\documentclass[preprint,%
secnumarabic,%
tightenlines,%
superscriptaddress,%
floats,prd,eqsecnum,nobibnotes%
,nofootinbib,aps,11pt,letterpaper]{revtex4}

\usepackage{palatino}
\usepackage{amsmath, amssymb}
\usepackage{subfigure, graphicx}
\usepackage[breaklinks=true,%
            plainpages=false,%
            bookmarks=true
           ]{hyperref}
\usepackage{color} 
\usepackage{braket} 
\usepackage{cancel}

\allowdisplaybreaks[3]
\newcommand\oldappendix{This will give an error if \backspace oldappendix
   already exists.}
\let\oldappendix\appendix
\renewcommand\appendix{\oldappendix%
   \renewcommand\theequation{\thesection.\arabic{equation}}
   \renewcommand\thesubsection{\thesection.\arabic{subsection}}}

\newcommand\be{\begin{equation}}
\newcommand\bea{\begin{eqnarray}}
\newcommand\ee{\end{equation}}
\newcommand\eea{\end{eqnarray}}

\newcommand\Regge{{\alpha^{\prime}}}
\newcommand{\bdm}{\begin{displaymath}}
\newcommand{\edm}{\end{displaymath}}

\newcommand{\f}[2]{\frac{#1}{#2}}
\newcommand{\bref}[1]{(\ref{#1})}

\renewcommand{\choose}[2]{\binom{#1}{#2}}

\newcommand\drm{\mathrm{d}}
\newcommand{\der}[2]{\frac{\drm #1}{\drm #2}}

\newcommand{\pd}{\partial}
\newcommand{\pdb}{\bar{\partial}}

\newcommand{\co}{\mathbb{C}}
\newcommand{\ints}{\mathbb{Z}}
\newcommand{\veps}{\varepsilon} 

\newcommand{\sech}{\text{\sech}}

\newcommand{\vev}[1]{\langle {#1}\rangle}

\newcommand{\vac}{\ket{\varnothing}}

\newcommand{\dg}{\dagger}
\newcommand{\id}{\openone}

\newcommand\normalTag{\addtocounter{equation}{1}\tag{\theequation}}

\makeatletter 

\newenvironment{calc}{
	\start@align\@ne\st@rredtrue\m@ne}
		     {\normalTag\endalign}

\makeatother 

\begin{document}

\title{\vspace*{\fill}Emission from the D1D5 CFT: Higher Twists}
\author{\large Steven G. Avery}\email{avery@mps.ohio-state.edu}
\affiliation{Department of Physics \\
      The Ohio State University \\
      191 West Woodruff Avenue \\
      Columbus, Ohio \ 43210-1117 \\ U.S.A.
      \vspace*{\fill}}
\author{\large Borun D. Chowdhury}\email{borundev@mps.ohio-state.edu}
\affiliation{Department of Theoretical Physics \\
Tata Institute of Fundamental Research \\
Homi Bhabha Road \\
Mumbai 400005 \\ India
	\vspace*{\fill}}

\begin{abstract}
\vspace*{\baselineskip}

We study a certain class of nonextremal D1D5 geometries and their
ergoregion emission. Using a detailed CFT computation and the
formalism developed in~\cite{acm1}, we compute the full spectrum and
rate of emission from the geometries and find exact agreement with the
gravity answer. Previously, only part of the spectrum had been
reproduced using a CFT description. We close with a discussion of the
context and significance of the calculation.

\vspace*{\fill}
\end{abstract}

\maketitle

\section{Introduction}\label{sec:intro}

The background geometry created by certain stacks of branes can be
broken into different regions. At asymptotic infinity, the geometry is
flat. As one moves radially inward one comes into a ``neck'' region,
which transitions between the flat space and an AdS ``throat''. The
AdS throat terminates in a ``fuzzball''
cap~\cite{BalaDeBoer,lm4,lm5,lmm,mss,gms1,gms2,st-1,st-2,st-3,st-4,bena-6,bena-7,bena-8,bena-9,ross,gimon-1,gimon-2,gimon-3,gimon-4,BDSMB,BSMB,fuzzballs2-1,fuzzballs2-2,fuzzballs2-3,fuzzballs2-4,giusto-1,giusto-2,giusto-3,giusto-4,giusto-5};
there are a large number of different caps with different structure.
These geometries with different caps are interpreted as the
microstates of the black hole formed from the branes. In general one
expects the microstates to be quite quantum; however, in many cases
one can find sets of classical supergravity solutions with different
caps, which can be used to gain intuition about the structure of these
microstates.

In~\cite{acm1}, a general formalism was developed to compute the
emission rate from the AdS region out into the asymptotically flat
region using the AdS/CFT correspondence. In the low energy limit,
modes living in the AdS region decouple from modes in the
asymptotically flat region, and the physics of the AdS region is
\emph{dual} to the physics of an appropriate CFT~\cite{gkp, witten,
Aharony}. One can consider, however, relaxing the strict decoupling
limit and allowing some small amount of interaction between the AdS
region and the flat region. Using the CFT description of the AdS
physics, the interaction is described by adding a new term to the
previously independent CFT and flat space actions, which couples flat
space fields to vertex operators of the CFT. The formalism was
demonstrated by computing the emission of minimal scalars from a
subset of D1D5 states found in~\cite{ross} and found exact agreement
with the gravity calculation.

In this paper, we extend the results of~\cite{acm1} by computing the
emission from a broader class of D1D5 states. In~\cite{cm3}, the
supergravity spectrum and rate of emission from this class of D1D5
geometries was \emph{partially} reproduced by doing a heuristic CFT
calculation. Our goal is to compute the rate of emission and
\emph{full} spectrum as a rigorous CFT calculation. This calculation
serves to demonstrate the method of~\cite{acm1} in a more nontrivial
calculation, and to tie up the loose ends of~\cite{cm3}.

The D1D5 system we work with lives in a ten-dimensional geometry
compactified on $T^4\times S^1$. We wrap $N_5$ D5 branes on the full
compact space, $T^4\times S^1$; and we wrap $N_1$ D1 branes on the
circle, $S^1$. This gives rise to an $AdS_3\times S^3$ throat in the
noncompact space, which is dual to a two-dimensional
CFT~\cite{maldacena,swD1D5, fmD1D5,deBoerD1D5,
dijkgraafD1D5,frolov-1,frolov-2, Jevicki,David}. The core AdS region
has radius $(Q_1Q_5)^\frac{1}{4}$, where the D1 and D5 charges $Q_1$
and $Q_5$ are given by
\begin{equation}
Q_1 = \frac{g\Regge^3}{V}N_1 \qquad Q_5 = g\Regge N_5,
\end{equation}
with  string coupling $g$ and $T^4$ volume $(2\pi)^4V$.

In~\cite{acm1}, the emission rate of minimal scalars from this $CFT_2$
was found to be
\begin{equation}\label{eq:D1D5-decay-rate}
\der{\Gamma}{E} = 
	\frac{2\pi}{2^{2l+1}\,l!^2}\frac{(Q_1Q_5)^{l+1}}{R^{2l+3}}(E^2-\lambda^2)^{l+1}\,
	|\bra{0}\hat{\mathcal{V}}(0)\ket{1}_\text{unit}|^2\,
	\delta_{\lambda,\lambda_0}\delta(E - E_0).
\end{equation}
In the above equation, $2\pi R$ is the coordinate volume of the $S^1$;
$l$ is the angular momentum quantum number of the emitted minimal
scalar; $E$ and $\lambda$ are the energy and momentum of the scalar
along the $S^1$; and $\mathcal{V}$ is the normalized vertex operator
in the CFT that couples to the minimal scalar. The CFT amplitude is
computed with ``unit'' $S^1$ coordinate $\sigma\in(0,2\pi)$ and
dimensionless Euclidean time $\tau = it/R$. The $(\sigma, \tau)$
dependence of the CFT amplitude has been integrated to give the
energy-momentum conserving delta-functions, and then removed from the
above equation by placing the vertex operator at the ``origin''. The
CFT states $\ket{1}$ and $\ket{0}$ are repectively the initial excited
state and the final state of the CFT. They are dual to the initial and
final states of the geometry's capped AdS region.

The vertex operator for emission of supergravity minimal scalars is
given by~\cite{acm1}
\begin{equation}\label{eq:vertex-op}
\widetilde{\mathcal{V}}^{A\dot{A}B\dot{B}}_{l, l-k-\bar{k}, k-\bar{k}}(\sigma, \tau)
	= \frac{1}{2}\sqrt{\frac{(l-k)!(l-\bar{k})!}{(l+1)^2(l+1)!^2\,k!\,\bar{k}!}}
	(J^+_0)^k(\bar{J}^+_0)^{\bar{k}}
	G^{+A}_{-\frac{1}{2}}\psi^{-\dot{A}}_{-\frac{1}{2}}
	\bar{G}^{\dot{+}B}_{-\frac{1}{2}}\bar{\psi}^{\dot{-}\dot{B}}_{-\frac{1}{2}}
	\tilde{\sigma}^0_{l+1}(\sigma, \tau).
\end{equation}
The subscript on $\widetilde{\mathcal{V}}_{l,m_\psi, m_\phi}$ is its charge
under the $SO(4)_E$ rotational symmetry of the $S^3$. The numerical
prefactor normalizes the operator so that it has unit two-point function
with itself at unit separation in the complex plane. Mapping to the
complex plane via
\begin{equation}\label{eq:map-to-the-plane}
z = e^{\tau + i \sigma}\qquad \bar{z} = e^{\tau - i\sigma},
\end{equation}
the vertex operator becomes
\begin{equation}\begin{split}
\widetilde{\mathcal{V}}^{A\dot{A}B\dot{B}}_{l, l-k-\bar{k}, k-\bar{k}}(\sigma, \tau) 
	&= |z|^{l+2}\mathcal{V}^{A\dot{A}B\dot{B}}_{l, l-k-\bar{k}, k-\bar{k}}(z, \bar{z}) \\
&= |z|^{l+2}
\frac{1}{2}\sqrt{\frac{(l-k)!(l-\bar{k})!}{(l+1)^2(l+1)!^2\,k!\,\bar{k}!}}
	(J^+_0)^k(\bar{J}^+_0)^{\bar{k}}
	G^{+A}_{-\frac{1}{2}}\psi^{-\dot{A}}_{-\frac{1}{2}}
	\bar{G}^{\dot{+}B}_{-\frac{1}{2}}\bar{\psi}^{\dot{-}\dot{B}}_{-\frac{1}{2}}
	\tilde{\sigma}^0_{l+1}(z, \bar{z}).
\end{split}\end{equation}
For more details of the CFT notation, consult the appendix A
of~\cite{acm1}.

The remainder of the paper may be broken into the following steps:
\begin{enumerate}
\item In Section~\ref{sec:physical-amp}, we set up the physical emission problem 
from the geometries of~\cite{ross}, in the CFT language developed
in~\cite{acm1}. Specifically, we describe the initial excited state
and the final state that go into the CFT amplitude, which when plugged
into Equation~\eqref{eq:D1D5-decay-rate} give the emission rate of the
minimal scalars. Evaluating the CFT amplitude is a nontrivial
exercise, to which the majority of the paper is dedicated. The initial
state is parameterized by three integers $n$, $\bar{n}$, and
$\kappa$. The positive integer $\kappa$, called $k$ in~\cite{cm3},
controls a conical defect  in the AdS region. The spectrum and rate of emission found
in~\cite{acm1} were for $\kappa=1$. In this paper, we extend those
calculations to $\kappa>1$.

\item We do not directly compute the physical CFT amplitude of interest. Instead, we 
compute a CFT amplitude that does \emph{not} correspond to a physical
gravitational process, and then map this ''unphysical'' amplitude onto
the physical problem using spectral flow and hermitian
conjugation. This route avoids some subtleties and allows us to use
some results from~\cite{acm1} in the calculation. In
Section~\ref{sec:unphysical-amp}, we precisely set up the unphysical
CFT amplitude to be computed compute.

\item In Section~\ref{sec:specflow-hermconj}, we show how to relate the 
unphysical amplitude to be computed to the physical problem using
spectral flow and Hermitian conjugation.

\item Before starting the lengthy calculation of the unphysical CFT amplitude, 
in Section~\ref{sec:method} we explain the specific method used to
compute it . The unphysical amplitude, we explain, can be lifted to a
covering space, where it becomes an amplitude computed
in~\cite{acm1}. The Jacobian factors that arise in mapping to the
covering space, however, are highly nontrivial. Additionally there are
some important combinatoric factors, which come in from symmetrizing
over all $N_1N_5$ copies of the CFT.

\item In Section~\ref{sec:T}, we compute the Jacobian factors that arise from 
mapping the twist operators to the covering space. We use the methods
for evaluating correlation functions of twist operators developed in~\cite{lm1}.

\item In Section~\ref{sec:M}, we compute the Jacobian factors produced by 
the non-twist operator insertions in the unphysical amplitude.

\item In Section~\ref{sec:comb}, we compute the combinatoric factors that come 
from symmetrizing over all $N_1N_5$ copies of the CFT. The result
simplifies in the large $N_1N_5$ limit that is physically relevant.

\item Finally in Section~\ref{sec:rate}, we use Section~\ref{sec:specflow-hermconj} 
to relate the computed unphysical amplitude to the final amplitude for
emission. We then plug the amplitude into
Equation~\eqref{eq:D1D5-decay-rate} to find the rate of emission. The
explicit $\kappa$-dependence comes in the form of a power,
$\kappa^{-2l-3}$, multiplying the rate for $\kappa=1$; the spectrum is
also affected. The spectrum and rate exactly match the gravity
calculation in~\cite{cm3}.
\end{enumerate}
Because this paper largely is a direct extension of~\cite{acm1}, we do
not introduce the notation used here in detail and instead refer the
reader to Appendix A of~\cite{acm1} for an overview of the CFT
notation. Furthermore, we take several results directly
from~\cite{acm1}.

\section{Emission from $\kappa$-orbifolded geometries}\label{sec:physical-amp}

For a fixed amount of D1 charge, D5 charge, and $S^1$-momentum,
~\cite{ross} found a three-parameter family of geometries that have
the following properties. At infinity they are asymptotically flat,
then as one goes radially inward one encounters a ``neck''
region. After passing through the neck, one finds an AdS throat which
terminates in an ergoregion cap. The fuzzball proposal interprets
these smooth, horizonless geometries as classical approximations to
microstates of a black hole with the same charges.

\begin{figure}[ht]
\begin{center}
\subfigure[~The gravity description.]{
\includegraphics[width=6cm]{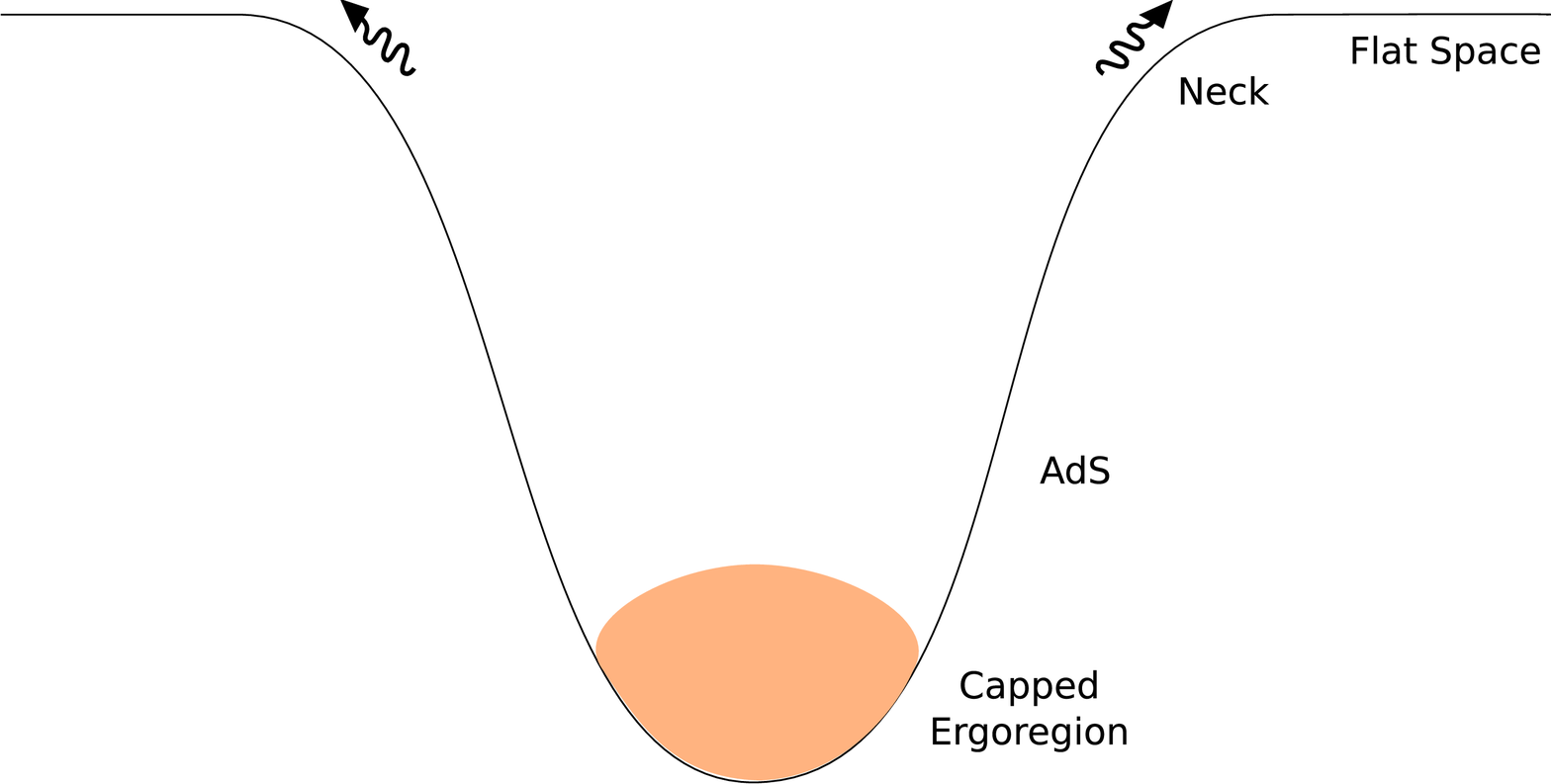}}
\hspace{30pt}
\subfigure[~The CFT description.]{
\includegraphics[width=6cm]{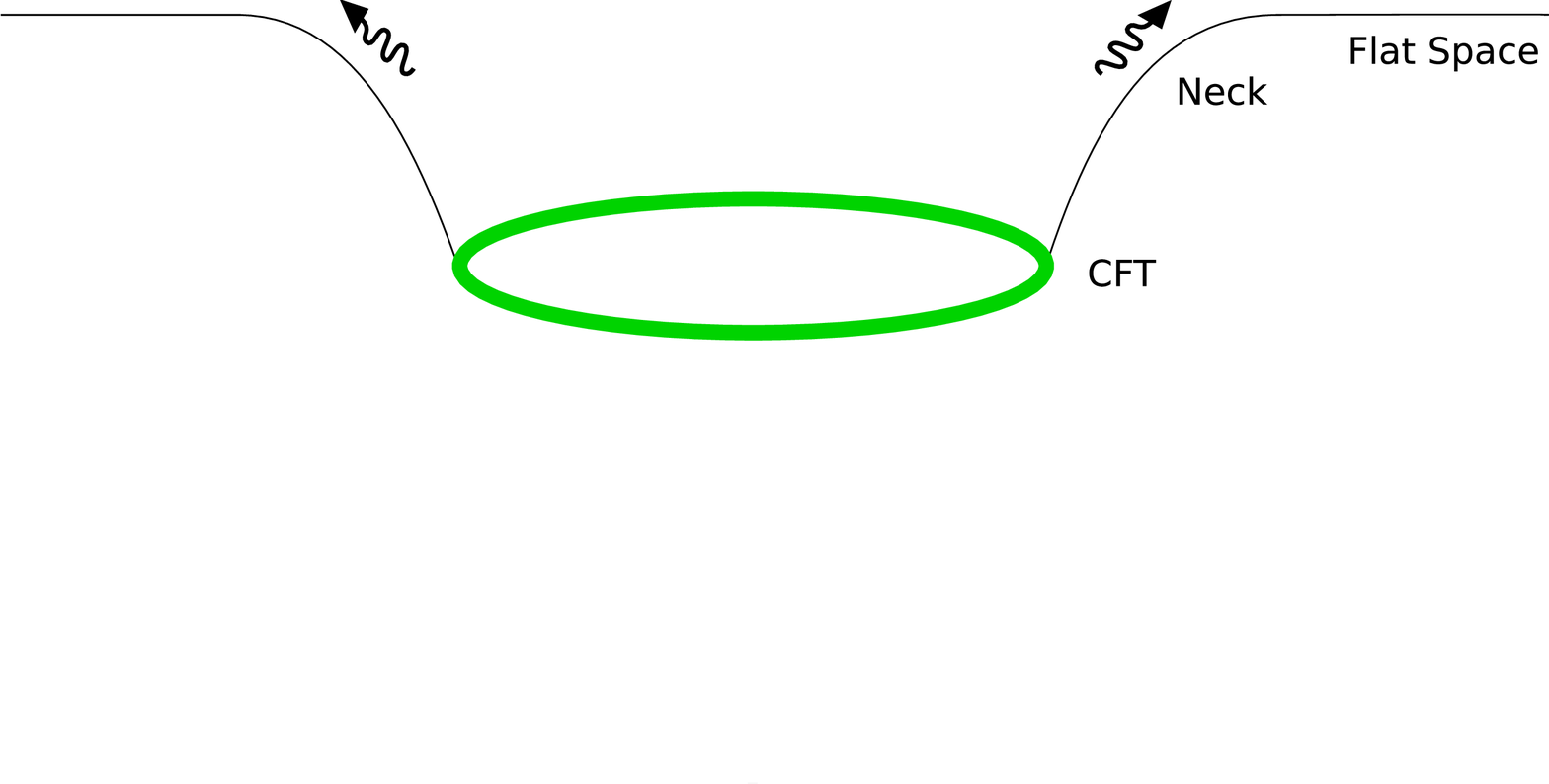}}
\caption{A depiction of the emission process we consider. In the cap 
region of the geometry, there is an ergoregion (shown in orange),
which leads to the emission of particles into the flat space. In the
CFT description particles are emitted by the CFT (shown in green)
directly into the ``neck'' region of the
geometry.}\label{fig:ergo-emission}
\end{center}
\end{figure}

The presence of ergoregions renders these geometries unstable
~\cite{myers,cm1,cm3}. The instability is exhibited by emission of
particles at infinity, with exponentially increasing flux, carrying
energy and angular momentum out of the geometry. In~\cite{cm1, cm2,
cm3}, using heuristic CFT computations for some special cases, it was
argued that the ergoregion emission is the Hawking radiation from this
subset of microstates for the D1D5 black
hole. Figure~\ref{fig:ergo-emission} depicts the emission process in
the gravity and CFT descriptions.

These geometries are dual to CFT states parameterized by three
integers $n$, $\bar{n}$, and $\kappa$. In~\cite{acm1}, the spectrum
and rate of emission from the geometries with $\kappa=1$ was exactly
reproduced with a CFT computation. In this paper, we calculate the
spectrum and rate of emission for $\kappa>1$. The parameter $\kappa$
(called $k$ in~\cite{cm3}), controls a conical defect in the
geometry. For $\kappa=1$, there is no conical defect or orbifold
singularity. 

Below, we first describe the initial state of the physical CFT
problem, which is dual to the unperturbed background geometry. Then,
we roughly describe the final state of the physical CFT amplitude that
the initial state decays to. We do not precisely give these states
since we do not directly compute with them. In
Section~\ref{sec:unphysical-amp}, we give the precise states used in
the ``unphysical'' CFT amplitude, which in
Section~\ref{sec:specflow-hermconj} we relate to the physically
relevant states of this section.

\subsection{The CFT initial state}

The physical initial state is the background geometry described
in~\cite{ross} and~\cite{cm3}. The decoupled AdS-part of the geometry
can be obtained by spectral flowing $\kappa$-orbifolded $AdS_3 \times
S^3$ by $\alpha = 2n+ \f{1}{\kappa}$ units on the left-moving sector
and by $\bar{\alpha} = 2\bar{n}+ \f{1}{\kappa}$ units on the
right-moving sector. The fractional spectral flow may seem strange;
however, it arises because of the conical defect. In
~\cite{mm,bal,gms2} geometries were constructed which had a decoupled
AdS part which can be understood in this context as spectral flow by
$\alpha=2n+\f{1}{\kappa}$ and $\bar \alpha=\f{1}{\kappa}$. However
these geometries are BPS and do not have an ergoregion and thus do not
radiate.  We discuss fractional spectral flow in the CFT context in
Section~\ref{sec:specflow}.

The $\kappa$-orbifolded $AdS_3$ is described, after a left and right
spectral flow by $\f{1}{\kappa}$, as twisting the $N_1N_5$ strands
into $\kappa$-length component strings in the R sector. This geometry
is stable and does not emit anything. Performing further spectral
flow, however, adds fermionic excitations to all of the component
strings and allows for the possibility of emission.

\begin{figure}[ht]
\begin{center}
\subfigure[~The physical initial state.]{\label{phy-initial}
\includegraphics[width=6cm]{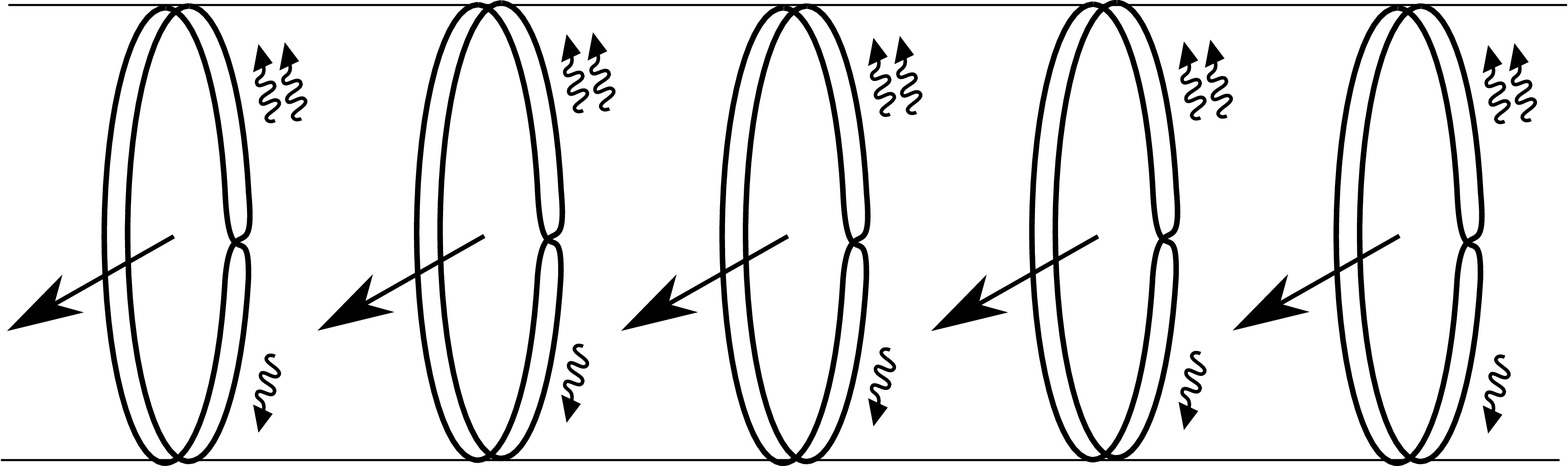}}
\hspace{30pt}
\subfigure[~The physical final state.]{\label{phy-final}
\includegraphics[width=6cm]{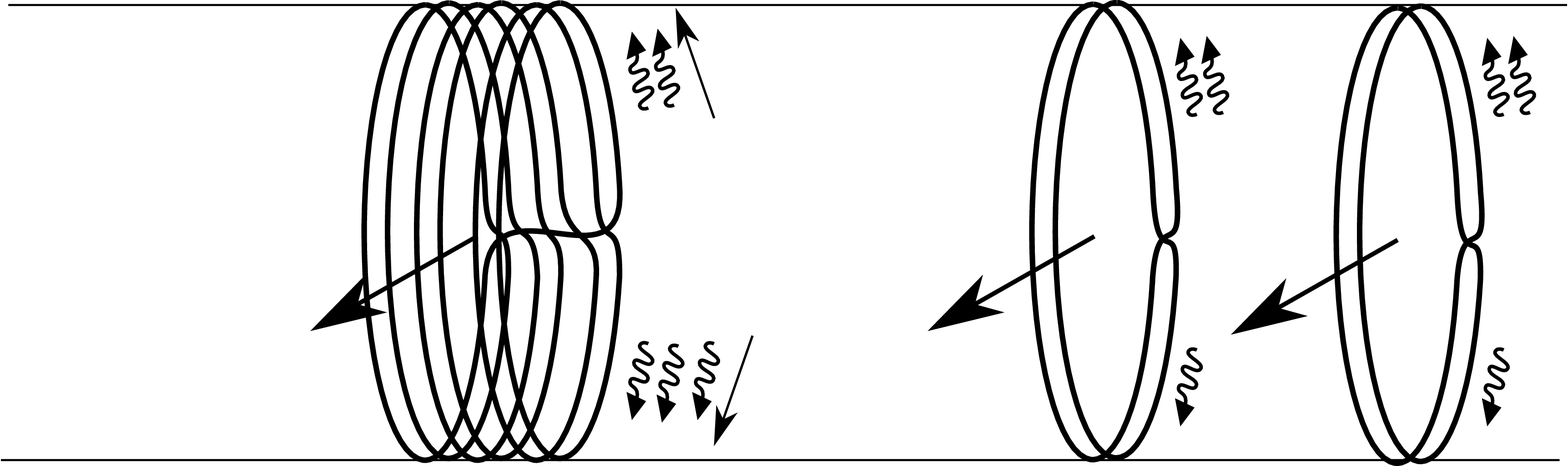}}
\caption{A depiction of the initial and final states of the 
physical amplitude.}\label{fig:physical-states}
\end{center}
\end{figure}

The $\kappa$-twisted component string in the Ramond vacuum has weight
$\kappa/4$. The Ramond vacuum also has ``base spin'' coming from the
fermion zero modes. Let us start with the ``spin up'' $\kappa$-twisted
Ramond vacuum with weight and charge
\begin{equation}
h = \bar{h} = \frac{\kappa}{4} \qquad m = \bar{m} = \frac{1}{2}.
\end{equation}
We then add energy and charge to this state by spectral-flowing by
$2n$ units, where spectral flow by $\alpha$ units affects the weight
and charge of a state by~\cite{spectral}
\begin{equation}\begin{split}
h &\mapsto h + \alpha m + \frac{\alpha^2 c_\text{tot.}}{24}\\
m &\mapsto m + \frac{\alpha c_\text{tot.}}{12}.
\end{split}\end{equation}
Spectral flowing the $\kappa$-twisted Ramond vacuum by $2n$ units
corresponds to filling all fermion energy levels up to the ``Fermi sea
level'' $n \kappa$. Keep in mind that there are two complex fermions
which have spin up and spectral flow fills fermion levels with both
fermions. This gives the initial state's weight and charge~\cite{cm3}
\begin{equation}
h_i = \kappa\left(n^2 + \frac{n}{\kappa} + \frac{1}{4}\right) 
	\qquad m_i = n\kappa + \frac{1}{2}
\end{equation}
and similarly for the right sector replacing $n$ by $\bar{n}$. This
state is depicted in Figure~\ref{phy-initial}.

\subsection{The CFT final state}

In emitting a particle of angular momentum $(l, m_\psi, m_\phi)$ the
initial state is acted on by the supergravity vertex operator
$\mathcal{V}_{l, -m_\psi, -m_\phi}$ from
Equation~\eqref{eq:vertex-op}.

We consider the process where the $(l+1)$-twist operator acts on $l+1$
distinct component strings of the background forming one long
$\kappa(l+1)$-twisted component string. The final state, then, is in
the $\kappa(l+1)$-twisted R sector. In principle there are other ways
that the vertex operator may twist the initial state; however, we work
in the limit where $N_1N_5\gg \kappa$, in which case this process
dominates. Processes in which the vertex operator twists more than one
strand of the same $\kappa$-length component string have probabilities
suppressed by factors of $1/(N_1N_5)$.

The charge/angular momentum of the vertex operator
$\mathcal{V}_{l,l-k-\bar{k}, k-\bar{k}}$ is given by
\begin{equation}\begin{split}
m_v &= -\left(\tfrac{l}{2} -k\right)\hspace{50pt} 
	m^\text{vertex}_\psi = -m_\psi = l-k-\bar{k}\\
\bar{m}_v &= -\left(\tfrac{l}{2} - \bar{k}\right)\hspace{50pt} 
	m^\text{vertex}_\phi = -m_\phi = k-\bar{k}.
\end{split}\end{equation}
From the charge of the vertex operator and the initial state it is
easy to deduce the charge of the final state:
\begin{equation}
m_f = (l+1)\left(n\kappa + \tfrac{1}{2}\right) - \left(\tfrac{l}{2} - k\right)
  = (l+1)n\kappa + k + \frac{1}{2}.
\end{equation}

There can be different final states with the same charge corresponding
to different harmonics of the emitted quanta. Since we derive the
amplitude using spectral flowed states it is not important to write
down the weights of the states here. However they are found easily by
using the charge and weight of the spectral flowed states. The final
state, which we call $\ket{f}$, is shown in Figure~\ref{phy-final}.

\subsection{The physical CFT amplitude}

From the initial and final states, the CFT amplitude which we use to
find the emission spectrum and rate is given by
\begin{equation}\label{eq:cylinder-amp}
\mathcal{A}_{l, m_\psi, m_\phi} 
	= \bra{f}\widetilde{\mathcal{V}}_{l, m_\psi, m_\phi}(\sigma, \tau)\ket{i}.
\end{equation}
We prefer to calculate on the complex plane with $z$ coordinates
instead of on the cylinder. Mapping the vertex operator to the plane
using Equation~\eqref{eq:map-to-the-plane} gives a Jacobian factor,
$|z|^{l+2}$, from its conformal weight. Thus, the physical CFT
amplitude that we ultimately wish to find is written as
\begin{equation}
\mathcal{A}_{l,m_\psi, m_\phi}(z) 
	= |z|^{l+2}\bra{f}\mathcal{V}_{l,m_\psi, m_\phi}(z, \bar{z})\ket{i}.
\end{equation}
The majority of the paper is expended in calculating this
amplitude. The tilde on the vertex operator in
Equation~\eqref{eq:cylinder-amp} is to distinguish it from the
operator in the complex plane.

\section{The ``unphysical'' amplitude}\label{sec:unphysical-amp}

We do not directly compute the amplitude of interest with the initial
and final states described above. Instead we compute an amplitude,
described below, that is related to the physical problem by spectral
flow and Hermitian conjugation. In this other amplitude, the
calculation is simpler and we have a better understanding of what the
initial and final states should be. In~\cite{acm1}, the same technique
was used.

In this section we describe the initial and final states of the
unphysical amplitude, $\ket{i'}$ and $\ket{f'}$, that we use to
calculate
\begin{equation}
\mathcal{A}' = \bra{f'}\mathcal{V}(z)\ket{i'}.
\end{equation}
Later, we relate this process to the physical problem by spectral flow
and Hermitian conjugation. Before we can give the states, we must
describe the notation we use, specifically addressing the fermions'
periodicity.

\subsection{Fermion periodicities}

First, let us emphasize what the R and NS sectors \emph{mean} in the
twisted sector.  We define two parameters $\beta_+$ and $\beta_-$, which
specify the periodicity of the fermions in the theory:
\begin{equation}
\psi^{\pm \dot{A}}\big(ze^{2\pi i}\big) = e^{i\pi\beta_\pm}\psi^{\pm\dot{A}}(z).
\end{equation}
Obviously, $\beta_\pm$ are only defined modulo two under addition.

The R sector means $\beta_\pm\equiv 1$, whereas the NS sector means
$\beta_\pm\equiv 0$. Spectral flow by $\alpha$ units has the effect of
taking
\begin{equation}\label{eq:spectral-eta}
\beta_\pm \mapsto \beta_\pm \pm \alpha.
\end{equation}

In the $p$-twisted sector, the boundary conditions are
\begin{equation}
\psi^{\pm\dot{A}}_{(j)}\big(ze^{2\pi i}\big) = e^{i\pi \beta_\pm}\psi^{\pm\dot{A}}_{(j+1)}(z),
\end{equation}
where $(j)$ indexes the different copies of the target space. This
implies that
\begin{equation}\label{eq:base-twisted-eta}
\psi^{\pm\dot{A}}_{(j)}\big(z e^{2p\pi i}\big) = e^{ip\pi\beta_\pm}\psi^{\pm\dot{A}}_{(j)}(z).
\end{equation}
In the base space, over each point $z$, there are $p$ different copies
of each field. To calculate, it is convenient to map to a covering
space where these $p$ copies become a single-valued field. When one
goes to a covering space the periodicity of the total single-valued
field $\Psi(t)$, then is given by
\begin{equation}
\Psi^{\pm\dot{A}}\big(te^{2\pi i}\big) = \exp\left[i\big(p\beta_\pm + (p-1)\big)\pi\right]
		\Psi^{\pm\dot{A}}(t),
\end{equation}
where the extra factor of $(p-1)$ in the exponent comes from the
Jacobian of the weight-half fermion, under a map of the form $z\propto
t^p$. We label the periodicity in the cover by
\begin{equation}\label{eq:cover-eta}
\beta^{(\mathrm{cover})}_\pm = p(\beta_\pm+1) -1.
\end{equation}
In the untwisted sector, we use the natural definition of the
fermions, $\psi(z)$, as being periodic, without branch cut. Thus,
application of a spin field operator is necessary to give antiperiodic
boundary conditions. In the twisted sector, the periodicity of
fermions in the base space is neither ``naturally'' periodic nor
antiperiodic since there is a hole and branch cut from the twist
operator; however, in the covering space, the fields are, again,
naturally periodic.

We denote ``bare twists'' which insert the identity in the cover as
\begin{equation}
\sigma_p(z, \bar{z}) \xrightarrow[\quad\text{cover}\quad]{\text{to the}} \id (t,\bar{t})
\qquad h = \bar{h} = \Delta_p = \frac{c}{24}\left(p - \frac{1}{p}\right).
\end{equation}
After the above discussion, we see that one should always use
\begin{equation}
(\text{NS Sector}) \,\Longrightarrow \, 
\begin{cases}
\sigma_p(z,\bar{z}) & \text{$p$ odd}\\
S_p^\alpha(z)\bar{S}_p^{\dot{\alpha}}(\bar{z})\sigma_p(z,\bar{z}) & \text{$p$ even}
\end{cases}, \qquad
(\text{R Sector}) \,\Longrightarrow \, 
S_p^\alpha(z)\bar{S}_p^{\dot{\alpha}}(\bar{z})\sigma_p(z, \bar{z}).
\end{equation}
The $S_p$'s indicate that one should insert a spin field in the
$p$-fold covering space at the image of the point $z$. In the above
statement, one could equally well choose $SU(2)_2$ indices instead of
$SU(2)_{L/R}$ indices for the spin fields.

The CFT amplitude we compute uses the bare twists, which for odd twist
order correspond to the NS sector and for even twist order correpond
to neither the NS nor the R sector. Even though this amplitude is not
physically relevant, we can use spectral flow to relate it to the R
sector process of physical interest. Below, we give the initial and
final states of the unphysical amplitude, starting with the simpler
final state.

\begin{figure}[ht]
\begin{center}
\subfigure[~The initial state of the unphysical amplitude.]
{\label{fig:unphy-in}
\includegraphics[width=6cm]{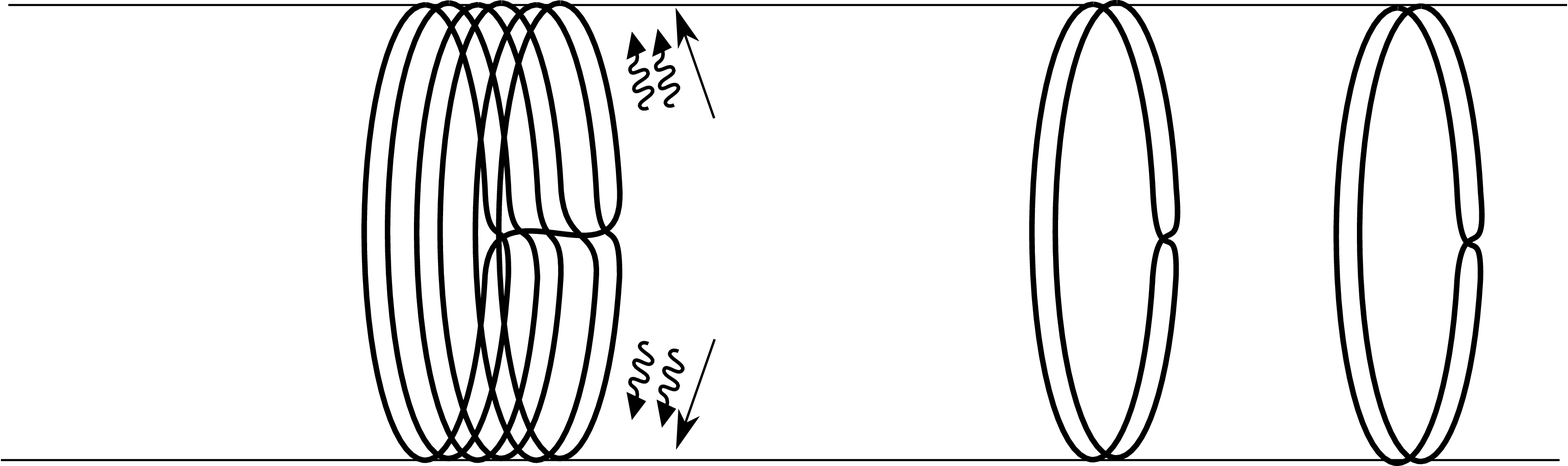}}
\hspace{30pt}
\subfigure[~The final state of the unphysical amplitude.]
{\label{fig:unphy-fin}
\includegraphics[width=6cm]{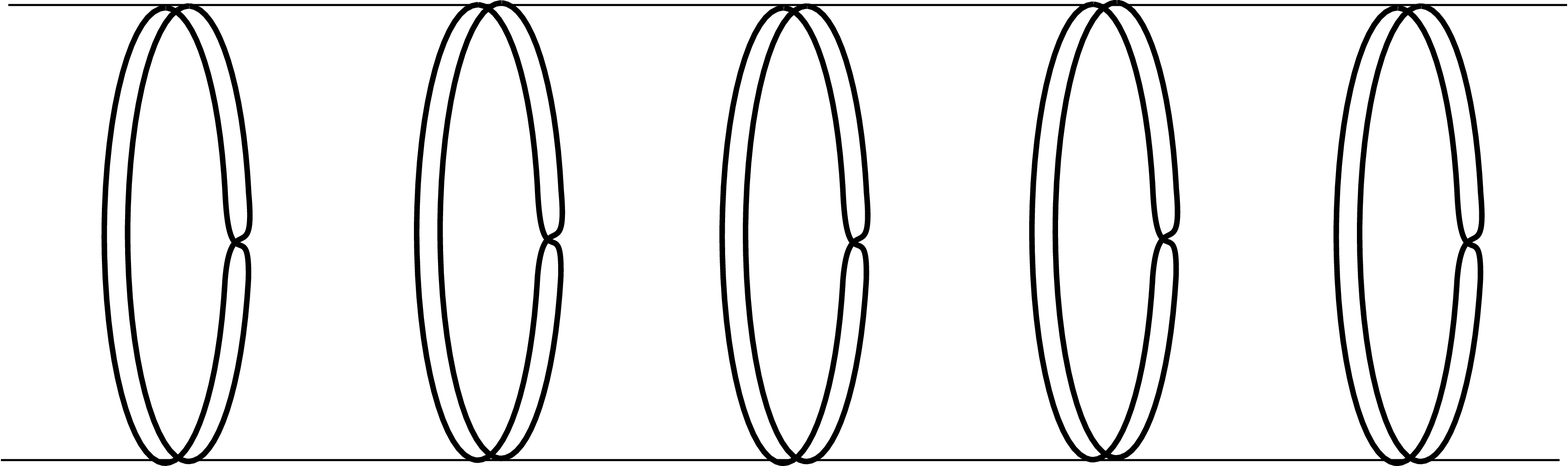}}
\caption{The initial and final states of the unphysical amplitude. 
The initial state spectral flows to the physical final state, and the
final state spectral flows to the physical initial
state.}\label{fig:unphysical-states}
\end{center}
\end{figure}

\subsection{The final state}

The \emph{final} state\footnote{Here, we just discuss the
$\kappa(l+1)$ strands that are involved in the nontrivial part of the
correlator. Later, we introduce the combinatoric factors which result
from symmetrizing over all $N_1N_5$ copies of the CFT.} spectral flows
to the \emph{initial} state of the physical problem. All of the
excitations and spin fields of the physical state can be acquired via
spectral flow, so we use bare twists for the unphysical amplitude
final state:
\begin{equation}
\ket{f'} = \underbrace{\sigma_\kappa\sigma_\kappa\cdots\sigma_\kappa}_{l+1}\vac_{NS}.
\end{equation}
This state has weight and charge
\begin{equation}
h_f = \bar{h}_f = \frac{1}{4}(l+1)\left(\kappa - \frac{1}{\kappa}\right)\qquad
m_f = \bar{m}_f = 0.
\end{equation}
Note that for odd $\kappa$ this state corresponds to the NS sector,
but for $\kappa$ even it corresponds to neither the NS nor the R
sector. This state, which we call $\ket{i}$, is shown in
Figure~\ref{fig:unphy-fin}.

\subsection{The initial state}

The \emph{initial} state spectral flows to the physical final state,
and therefore consists of excitations in the $\kappa(l+1)$-twisted
sector, as shown in Figure~\ref{fig:unphy-in}. Below, modulo
normalization, we give the left part of the state
\begin{equation}
\ket{i'} = 
\big(J_0^-\big)^k
G^{- A}_{-\frac{1}{2\kappa}}\big(L_{-\frac{1}{\kappa}}\big)^N \psi^{+\dot{A}}_{-\frac{1}{2\kappa}}
\begin{cases}
	J^+_{-\frac{l-1}{\kappa(l+1)}}J^+_{-\frac{l-3}{\kappa(l+1)}}
		\cdots
	J^+_{-\frac{1}{\kappa(l+1)}} \sigma_{\kappa(l+1)}\vac_{NS} &
		\text{$(l+1)$ odd}\\
		J^+_{-\frac{l-1}{\kappa(l+1)}}J^+_{-\frac{l-3}{\kappa(l+1)}}
			\cdots
		J^+_{-\frac{2}{\kappa(l+1)}}
	S^+_{\kappa(l+1)}\sigma_{\kappa(l+1)}\vac_{NS}  &  \text{$(l+1)$ even}.
\end{cases}
\end{equation}
The normalization of the state is $\kappa$-dependent, which plays an
important role in the calculation as discussed in
Section~\ref{sec:Mi}.

This state has weight and charge
\begin{equation}
h_i = \frac{1}{\kappa}\left(\frac{l}{2} + N + 1\right) 
		+ \frac{1}{4}(l+1)\left(\kappa - \frac{1}{\kappa}\right)\qquad
m_i = \frac{l}{2} - k,
\end{equation}
and similarly for the right sector. We conjecture the above form for
the initial state based on the $\kappa=1$ case in~\cite{acm1} and work
done in~\cite{cm3} for $\kappa>1$. The excited initial state should be
dual to a supergravity excitation of the orbifolded-AdS background. In
global AdS, one identifies the supergravity duals as the descendants
of chiral primary states under the anomaly-free subalgebra; here, we
propose the above modification for the $\kappa$-orbifolded
background. This is motivated, in part, by the action of the vertex
operator for absorption of supergravity particles on
$\kappa$-orbifolded AdS.

\subsection{The amplitude}

The amplitude we compute, then, is
\begin{equation}
{\mathcal{A}'}^\kappa = \bra{f'}\mathcal{V}_{l, l- k-\bar{k}, k-\bar{k}}(z_2)\ket{i'}.
\end{equation}
Note that if we map to the $\kappa$-cover, then these states are the
initial and final states of the $\kappa=1$ calculation in~\cite{acm1}.
In mapping to a $\kappa$-covering space the final state's twist
operators are removed, and the initial state's twist operator becomes
an $(l+1)$-twist. The $J^+$ modes acting on the initial state, along
with the spin field for even $l+1$, act on the $(l+1)$-twist in the
$\kappa$-cover in precisely the correct way to form the lowest weight
chiral primary twist operator of~\cite{lm2}:
\begin{multline}
J^+_{-\frac{l-1}{\kappa(l+1)}}J^+_{-\frac{l-3}{\kappa(l+1)}}
		\cdots
\begin{cases}
	J^+_{-\frac{1}{\kappa(l+1)}} \sigma_{\kappa(l+1)}\\
		J^+_{-\frac{2}{\kappa(l+1)}}
	S^+_{\kappa(l+1)}\sigma_{\kappa(l+1)}
\end{cases}\hspace{-15pt}
 \xrightarrow[\text{$\kappa$-cover}]{\text{to}}
J^+_{-\frac{l-1}{l+1}}J^+_{-\frac{l-3}{l+1}}\cdots
\begin{cases}
	J^+_{-\frac{1}{l+1}} \sigma_{l+1}\\
	J^+_{-\frac{2}{l+1}}
	S^+_{l+1}\sigma_{l+1}\\
\end{cases}\hspace{-15pt}
= \sigma^0_{l+1}.
\end{multline}
We follow the notation of~\cite{acm1} by denoting the lowest weight
chiral primary ${(l+1)}$-twist operator, introduced in~\cite{lm2}, as
$\sigma^0_{l+1}$. The above equation does not include Jacobian
factors, computed in Sections~\ref{sec:T} and~\ref{sec:M}, which are
introduced in going to the $\kappa$-covering space.

The specific covering space to which we map preserves the form of the
$(l+1)$-twist vertex operator. The fact that going to a $\kappa$
covering space gives the $\kappa=1$ amplitude along with the gravity
description of the emission process, is precisely the motivation for
introducing the specific form of the state
$\ket{i'}$. 

\section{Relating the computed CFT amplitude to the physical problem}\label{sec:specflow-hermconj}

There are two steps needed to map the physical problem onto the
unphysical CFT computation. First, we use spectral flow to map the
physical states, $\ket{i}$ and $\ket{f}$, to the ``primed states'',
$\ket{f'}$ and $\ket{i'}$.  Second, we use Hermitian conjugation to
reverse the initial and final primed states.

\subsection{Using spectral flow}\label{sec:specflow}

We wish to relate a CFT amplitude computed with the unphysical
``primed states'',
\begin{equation}
\mathcal{A}' = \bra{f'}\mathcal{V}(z, \bar{z})\ket{i'},
\end{equation}
to the physical amplitude in the Ramond sector.  In this section, we
show how to spectral flow~\cite{spectral,spectral-yu,vafa-warner} the
physical problem in the R sector to the actual CFT amplitude we
compute.

If spectral flowing the states $\ket{i'}$ and $\ket{f'}$ by $\alpha$
units is given by
\begin{equation}
\ket{f'}\mapsto \ket{i} = \mathcal{U}_\alpha\ket{f'}\qquad
\bra{i'}\mapsto \bra{f} = \bra{i'}\mathcal{U}_{-\alpha},
\end{equation}
then we can compute the physical Ramond sector amplitude for emission
of a particle with angular momentum $(l, m_\psi, m_\phi)$ by using
\begin{calc}
\mathcal{A}_{l, m_\psi, m_\phi} &= 
	|z|^{l+2}\bra{f} \mathcal{V}_{l, -m_\psi, -m_\phi}(z,\bar{z})\ket{i}\\
 	  &= |z|^{l+2}\big(\bra{f}\mathcal{U}_{\alpha}\big)
	     \big(\mathcal{U}_{-\alpha}\mathcal{V}_{l, -m_\psi, -m_\phi}\mathcal{U}_{\alpha}\big)
			     \big(\mathcal{U}_{-\alpha}\ket{i}\big)\\
	  &= |z|^{l+2}\bra{i'}\mathcal{V}'_{l, -m_\psi, -m_\phi}(z, \bar{z})\ket{f'}.
\end{calc}
Note that one finds $\mathcal{V}'$ by spectral flowing $\mathcal{V}$
by $-\alpha$ units.

In~\cite{acm1}, it was shown that the vertex operator transforms under
spectral flow as
\begin{equation}
\mathcal{V}'_{l,l-k-\bar{k}, k-\bar{k}}(z,\bar{z}) = z^{-\alpha(\frac{l}{2}-k)}
	\bar{z}^{-\bar{\alpha}(\frac{l}{2}-\bar{k})}
		\mathcal{V}_{l,l-k-\bar{k}, k-\bar{k}}(z,\bar{z}).
\end{equation}
We can use the vertex operator's transformation to write
\begin{equation}\label{eq:spectral-flow-A}
\mathcal{A}_{l, k+\bar{k} -l, \bar{k} - k} 
	= z^{-\alpha(\frac{l}{2}-k)}\bar{z}^{-\bar{\alpha}(\frac{l}{2}-\bar{k})}\,
	|z|^{l+2}\bra{i'}\mathcal{V}_{l,l-k-\bar{k}, k-\bar{k}}(z,\bar{z})\ket{f'}.
\end{equation}
Note that the primed states are reversed from what one would like.

The spectral flow parameter, $\alpha$, is chosen to have a value that
spectral flows the primed states to the physical states, which is
achieved by
\begin{equation}
\alpha = 2n+\frac{1}{\kappa}\qquad \bar{\alpha} = 2\bar{n} + \frac{1}{\kappa} 
	\qquad n,\bar{n}\in\ints.
\end{equation}
Spectral flowing by non-integer units may seem strange, however, one
can show that this results from the peculiar bare twist
operators. Consider spectral flowing the bare twist $\sigma_\kappa$ by
$\frac{1}{\kappa}$ units, and recall that under spectral flow by
$\alpha$ units
\begin{equation}
\beta_\pm \mapsto \beta_\pm \pm \alpha,
\end{equation}
and
\begin{equation}
\beta^{(\text{cover})}_\pm = \kappa(\beta_\pm + 1) - 1.
\end{equation}
The bare twist has $\beta_\pm^{(\text{cover})}=0$, which after
spectral flowing by $1/\kappa$ units becomes
${\beta_\pm^{(\text{cover})}=1}$ consistent with the R boundary
conditions in the base space. Similarly, one finds that that the
fermion periodicity of $\ket{i'}$ becomes correct for the R sector
after spectral flowing by $1/\kappa$ units. After spectral flowing by
$1/\kappa$ units to get to the R sector, we spectral flow by an even
number of units to build up the fermionic excitations of the state
$\ket{i}$.

\subsection{Hermitian conjugation}

Having related the physical amplitude to
\begin{equation}
\bra{i'}\mathcal{V}_{l, m_\psi, m_\phi}(z,\bar{z})\ket{f'},
\end{equation}
we now wish to switch the initial and final primed states by Hermitian
conjugation.

On the cylinder, the vertex operator Hermitian conjugates as
\begin{equation}
\left[\widetilde{\mathcal{V}}_{l, m_\psi, m_\phi}(\tau,\sigma)\right]^\dg 
	=  \widetilde{\mathcal{V}}_{l, -m_\psi, -m_\phi}(-\tau, \sigma),
\end{equation}
and therefore
\begin{equation}
\bra{i'}\widetilde{\mathcal{V}}_{l, m_\psi, m_\phi}(\tau,\sigma)\ket{f'}
 = \left[\bra{f'} \widetilde{\mathcal{V}}_{l,-m_\psi, -m_\phi}(-\tau, \sigma)\ket{i'}\right]^\dg.
\end{equation}
In the complex plane, this statement translates to
\begin{equation}
|z|^{l+2}\bra{i'}\mathcal{V}_{l, m_\psi, m_\phi}(z, \bar{z})\ket{f'}
  = |z|^{-(l+2)}\left[\bra{f'}
	\mathcal{V}_{l, -m_\psi, -m_\phi}\left(\tfrac{1}{z}, \tfrac{1}{\bar{z}}\right)
	\ket{i'}\right]^\dg.
\end{equation}

Applying this result to Equation~\eqref{eq:spectral-flow-A}, we may
write the physical CFT amplitude for emission of a particle with
angular momentum $(l, k+\bar{k}, \bar{k}-k)$ as
\begin{calc}\label{eq:unphysical-to-physical}
\mathcal{A}_{l, k+\bar{k}-l, \bar{k}-k} 
	&= z^{-\alpha(\frac{l}{2} - k)}\bar{z}^{-\bar{\alpha}(\frac{l}{2} - \bar{k})}
		|z|^{-(l+2)} 
	\left[\bra{f'}
	\mathcal{V}_{l, k+\bar{k}-l, \bar{k}-k}\left(\tfrac{1}{z}, \tfrac{1}{\bar{z}}\right)
	\ket{i'}\right]^\dg\\
	&= z^{-\frac{l}{2}-1-\alpha(\frac{l}{2} - k)}
		\bar{z}^{-\frac{l}{2} -1 -\bar{\alpha}(\frac{l}{2} - \bar{k})}
	\left[\mathcal{A}'_{l, k+\bar{k}-l, \bar{k}-k} 
		\left(\tfrac{1}{z}, \tfrac{1}{\bar{z}}\right)\right]^\dg.
\end{calc}
Thus, we compute the unphysical amplitude $\mathcal{A}'(z, \bar{z})$,
and use the above relation to find the physical amplitude
$\mathcal{A}$. The unphysical amplitude $\mathcal{A}'$ is independent
of $k$ and $\bar{k}$, so we suppress the subscripts.

\section{Method of computation}\label{sec:method}

The way we compute the CFT amplitude ${\mathcal{A}'}^\kappa$ is by noting
that if one were to lift to the $\kappa$-cover, the amplitude becomes
${\mathcal{A}'}^{\kappa=1}$ computed in~\cite{acm1}. Specifically, if we
transform to a coordinate $u$ via the map
\begin{equation}
z = b u^\kappa\qquad b = \frac{z_2}{u_2^\kappa},
\end{equation}
where $u_2$ is the image of the point $z_2$; then, the operators left
in the correlator are exactly those for ${\mathcal{A}'}^{\kappa=1}$.
In lifting to the cover, however, we do get some nontrivial ``Jacobian
factors'' from the various operators in the correlator transforming
under the map. Therefore, we write
\begin{equation}
{\mathcal{A}'}^\kappa 
	= \left(\frac{{\mathcal{A}'}^\kappa}{{\mathcal{A}'}^{\kappa=1}}\right)
		{\mathcal{A}'}^{\kappa=1}
	= TM{\mathcal{A}'}^{\kappa=1},
\end{equation}
where we have written the Jacobian factors as the product of two
contributions: one from the twists, $T$, and one from the modes, $M$.

\begin{figure}[ht]
\begin{center}
\includegraphics[width=4cm]{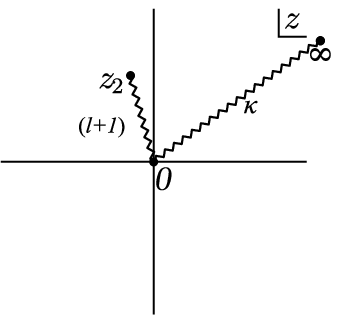}
\hspace{20pt}
\raisebox{1.8cm}{$\xrightarrow[]{\hspace{14pt}z\,=\,b u^\kappa\hspace{10pt}}$}
\hspace{20pt}
\includegraphics[width=4cm]{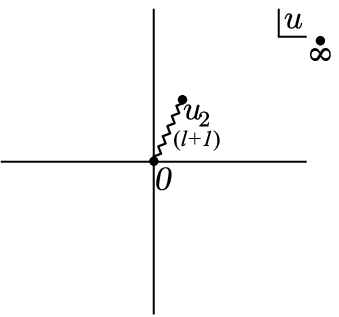}
\caption{In the $z$-plane, there is a twist operator of order
  $\kappa(l+1)$ at the origin, a twist of order $l+1$ at a point we
  call $z_2$, and $l+1$ $\kappa$-twists at infinity. The diagram on
  the left depicts the branch cuts connecting the three points. The
  number next to the branch cut depicts its ``order''. One can remove
  the $l+1$ branch cuts of order $\kappa$ between the origin and
  infinity by mapping to the $u$-coordinate, as shown in the figure on
  the right.}\label{fig:branch-cuts}
\end{center}
\end{figure}

We begin with the lengthier calculation of the nontrivial factor $T$
that comes from the twist operators transforming. One can see that $T$
is given by the ratio
\begin{equation}\label{eq:T-def}
T = \lim_{z_3\to\infty}
    |z_3|^{4(l+1)\Delta_\kappa}
  \frac{\vev{\sigma_\kappa\dots\sigma_\kappa(z_3)\sigma_{l+1}(z_2)\sigma_{\kappa(l+1)}(0)}}
	{\vev{\sigma_{l+1}(u_2)\sigma_{l+1}(0)}},
\end{equation}
where all of the above twists are ``bare'' twists which give \emph{no}
insertions in total $\kappa(l+1)$-covering space. The $z_3$ prefactor
ensures that in the limit as $z_3\to\infty$ the $l+1$ twists at
infinity make a normalized bra.

After computing $T$, we compute the mode Jacobian factor $M$, which is
given by the product of all of the Jacobian factors that the modes
circling the origin and the point $z_2$ acquire. 

\section{Computing the twist jacobian factor $T$}\label{sec:T}

Since the \emph{effect} of the twist operators is inherently nonlocal,
they do not transform in a simple local way when lifting to a covering
space. The simplest way to compute the numerator of
Equation~\eqref{eq:T-def}, is by lifting to the total covering space
and then computing the Liouville action that comes in conformally
transforming the induced metric on the covering space to a fiducial
metric. This is the method developed in~\cite{lm1}. Recently,
\cite{rastelli1, rastelli2} developed new technology for correlators
of $S_{N_1N_5}$-twist operators; however, we do not use those
techniques in this paper.

The denominator of Equation~\eqref{eq:T-def} is just the two-point
function of some normalized twist operators, and so we know the
answer:
\begin{equation}
\vev{\sigma_{l+1}(u_2)\sigma_{l+1}(0)} = \frac{1}{|u_2|^{\frac{c}{6}\left(l+1 - \frac{1}{l+1}\right)}}
	= \frac{1}{|u_2|^{4\Delta_{l+1}}}.
\end{equation}

From the weights of the twist operators and $SL(2,\co)$ invariance of
the correlator, we know that any three-point function of quasi-primary
fields must be given by
\begin{equation}\begin{split}
&\vev{\mathcal{O}_3(z_3)\,\mathcal{O}_2(z_2)\,
	\mathcal{O}_1(z_1)}\\
&\qquad= \frac{|C_{123}|^2}{
	\left(|z_1 - z_2|^{\Delta_1 + \Delta_2 - \Delta_3}
	       |z_2 - z_3|^{\Delta_2 + \Delta_3 - \Delta_1}
	       |z_3 - z_1|^{\Delta_3 + \Delta_1 - \Delta_2}\right)^2}.
\end{split}\end{equation}
In the present case where $z_3 = \infty$ and $z_1 = 0$, this gives
(after regularizing the divergence which corresponds to correctly
normalizing the final state)
\begin{equation}\label{eq:three-point-function}
\lim_{z_3\to\infty}|z_3|^{4(l+1)\Delta_\kappa}
\vev{\sigma_\kappa\dots\sigma_\kappa(z_3)\sigma_{l+1}(z_2)\sigma_{\kappa(l+1)}(0)}
 = \frac{|C|^2}{|z_2|^{2(\Delta_{\kappa(l+1)} + \Delta_{l+1} - (l+1)\Delta_\kappa)}}.
\end{equation}
It is the constant $C$ which is nontrivial and is determined in the
calculation that follows. From the above, one sees that the twist
Jacobian factor is completely determined by the constant $C$:
\begin{equation}\label{eq:C-gives-T}
T = |C|^2 \frac{|u_2|^{4\Delta_{l+1}}}
	{|z_2|^{2(\Delta_{\kappa(l+1)} + \Delta_{l+1} - (l+1)\Delta_\kappa)}}.
\end{equation}

\subsection{Review of method to compute correlators of twist operators}

We use the methods developed in~\cite{lm1} to compute the correlator
in Equation~\eqref{eq:three-point-function}. We call the physical
space where the problem is posed the ``base space'' which has the
coordinates $z$ and $\bar{z}$. In the base space, the basic fields are
multivalued. The basic method to computing the correlators of twist
operators consists of using a meromorphic mapping to a covering space
where there is one set of single-valued fields. Having single-valued
fields comes at the expense of introducing curvature in the covering
space. Fortunately, we can conformally map the curved covering space
to a manifold with metric in a fiducial (flat) form. Because of the
conformal curvature anomaly, the path integral is not invariant under
conformal mappings; however, the path integral changes in a specific
way.

Suppose we compute the path integral of our conformal field theory on
the manifold with fiducial metric $\widehat{ds}^2$, which we call
$\hat{Z}$. The path integral of the same conformal field theory on a
manifold with metric $ds^2 = e^{\phi} \widehat{ds}^2$ which we call
$Z$, is related to $\hat{Z}$ by~\cite{liouville}
\begin{equation}
Z = e^{S_L}\hat{Z},
\end{equation}
where the Liouville action, $S_L$, is given by
\begin{equation}
S_L = \frac{c}{96\pi}\int\drm^2 t\sqrt{-\hat{g}}\bigg[
	\hat{g}^{\mu\nu}\pd_\mu\phi\pd_\nu\phi + 2 \hat{R}\phi\bigg].
\end{equation}
$\hat{g}$ is the fiducial metric, and $\hat{R}$ is the Ricci
scalar of the \emph{fiducial} metric.

We now outline the precise steps needed to compute correlators of
twist operators, as given in~\cite{lm1}. The problem is posed in terms
of some twist operators inserted at various points $z_i$ in the base
space. To make the fields well-defined, we cut a small hole of radius
$\veps$ around each $z_i$ and demand that the fields have the correct
twisted boundary conditions around that hole. Furthermore, we regulate
the path integral by putting the correlator on a disc of radius
$1/\delta$, which encloses all of the $z_i$ except twist operators
that are inserted at infinity. To define the boundary conditions on
the edge of the disc, we glue a second flat disc onto the edge of the
first, giving the base space the topology of a sphere. Any twist
operators located at infinity, we insert at the center of this second
disc.  We work on the base space with metric
\begin{equation}\label{eq:base-metric}
ds^2 = \begin{cases}
dzd\bar{z}, & |z| < \frac{1}{\delta}\\
d\tilde{z}d\bar{\tilde{z}}, & |\tilde{z}| < \frac{1}{\delta}
\end{cases},\hspace{30pt}
\tilde{z} = \frac{1}{\delta^2}\frac{1}{z}.
\end{equation}
The path integral with the various regularizations on the above metric
we write as
\begin{equation}
Z^{(s)}_{\veps, \delta}[\sigma_{n_1}(z_1)\cdots\sigma_{n_N}(z_N)].
\end{equation}

We can map to a covering space with metric $ds'^2$ with coordinates $t$
and $\bar{t}$ via a meromorphic function $z = z(t)$. Then, the path
integral is given by
\begin{equation}\label{eq:corr-to-covering-space}
Z^{(s)}_{\veps,\delta}[\sigma_{n_1}(z_1)\cdots\sigma_{n_N}(z_N)] 
   = Z^{(s')}_{\veps,\delta},
\end{equation}
where
\begin{equation}
ds^2 = dzd\bar{z} \quad \Longrightarrow \quad 
	{ds'}^2 = \der{z}{t}\der{\bar{z}}{\bar{t}}\,dtd\bar{t}.
\end{equation}
When we write $Z^{(s')}$ with no square brackets we mean that there
are no insertions in the path integral. Throughout this discussion, we
assume that we are computing the correlator of bare twists which leave
no insertions in the covering space. If this were not the case, then
the right hand side of Equation~\eqref{eq:corr-to-covering-space} is
multiplied by the separately computed correlator of those insertions
in the covering space.

The sizes of the various holes in the $t$ coordinates are determined
by the mapping. The boundary conditions at the edges of the holes are
defined by pasting in flat pieces of manifold in the covering
space\footnote{If the twist operators are not bare twists, then there
would also be some operator insertions in the covering space; however,
here we just discuss bare twists.}.

The induced metric on the covering space, ${ds'}^2$, is conformally
related to the fiducial metric we choose to work with
\begin{equation}\label{eq:covering-metric}
\widehat{ds}^2 = \begin{cases}
dtd\bar{t}, & |t| <\frac{1}{\delta'}\\
d\tilde{t}d\bar{\tilde{t}}, & |\tilde{t}| < \frac{1}{\delta'}
\end{cases},\hspace{30pt}
\tilde{t} = \frac{1}{\delta'^2}\frac{1}{t}.
\end{equation}
Thus, we can write
\begin{equation}
Z_{\veps,\delta}^{(s)}[\sigma_{n_1}(z_1)\cdots\sigma_{n_N}(z_N)] 
    = e^{S_L}Z^{(\hat{s})}_{\veps,\delta, \delta'}.
\end{equation}
The correlator of the regularized twist operators is defined by
\begin{equation}\label{eq:def-corr}
\vev{\sigma_{n_1}(z_1)\cdots\sigma_{n_N}(z_N)}_{\veps, \delta} 
   = \frac{Z^{(s)}_{\veps,\delta}[\sigma_{n_1}(z_1)\cdots\sigma_{n_N}(z_N)]}
	{\big(Z_{\delta}\big)^{s}}
   = e^{S_L} \frac{Z_{\veps, \delta, \delta'}^{(\hat{s})}}{\big(Z_\delta\big)^s},
\end{equation}
where the $s$ in the denominator is the number of Riemann sheets or
the number of copies involved in the correlator (not to be confused
with the metric). The path integral in the numerator, then, is also
only over the twisted copies of the CFT, while the path integral with
no insertions in the denominator, $Z_\delta$, is the path integral of
a single copy of the CFT with the metric in
Equation~\eqref{eq:base-metric}.

The path integrals on the right-hand side of
Equation~\eqref{eq:def-corr} have no insertions and are on a metric of
identical form. For the case where the covering space manifold has
spherical topology, the path integrals cancel out up to the $\delta$
and $\delta'$ dependence.

To compute $C$ in Equation~\eqref{eq:three-point-function} and thereby
the twist Jacobian factor, $T$, we undertake the following steps:
\begin{enumerate}
\item We find the map to the total covering space, $z=z(t)$, that has the required
  properties.
\item We compute the Liouville action that comes from
  conformally mapping the induced metric of the total covering space
  to the fiducial form in Equation~\eqref{eq:covering-metric}.
\item The above correlators are all of normalized twist operators. The
  normalization is such that the two-point function has unit
  correlator at unit separation, as determined in~\cite{lm1}. We put
  in the normalization factors, which cancel out the
  $\veps$-dependence of the correlator.
\item Finally, we put all of the pieces together to determine $C$,
  and thereby $T$.
\end{enumerate}
Step 2 we relegate to Appendix~\ref{sec:liouville-action}.

\subsection{The map to the total covering space}

The first step in the calculation of $T$, then, is to find the map to
the total covering space. We have an $SL(2,\co)$ symmetry of the
covering space that allows us to fix three points. We fix the image of
the origin of the $z$-plane to the origin of the $t$-plane, the image
of $z_2$\footnote{We should point out that there are other images of
$z_2$, but only $t=1$ corresponds to where the vertex operator
acts. The fact that there are other images of $z_2$ corresponds to the
fact that the vertex operator only acts on $l+1$ strands of the
$\kappa(l+1)$ initial strands.} to the point $t=1$, and one image of
$z=\infty$ to $t=\infty$. Having expended the $SL(2,\co)$ symmetry, we
do not have freedom to fix the locations of the other images of
infinity in the covering space. We call the other images of infinity
$t_\infty^{(j)}$, where $j=1,\dots, l$.

Thus, we require our map to behave as
\begin{equation}\begin{aligned}
z &\sim t^{\kappa(l+1)}  & z&\approx 0,\; t\approx 0\\
z -z_2 &\sim  (t-1)^{l+1} &  z &\approx z_2,\; t\approx 1\\
z &\sim t^\kappa  & z &\to \infty,\; t\to \infty\\
z &\sim (t-t_\infty^{(j)})^{-\kappa} & z &\to\infty,\; t\to t_\infty^{(j)},
\end{aligned}\end{equation}
and be regular at all other points.  Here, $\sim$ means proportional
to at leading order.  Generically, one expects $\kappa(l+1)$ images of
infinity. However, in this case there should only be $l+1$ images of infinity since there
are only $l+1$ disconnected strings at $z\to\infty$. A priori, we do
not know the $t_\infty^{(j)}$; this is part of the output in finding
the map.

From the above conditions and requiring regularity everywhere but the
above special points, we see that $\der{z}{t}$ may have zeroes only at
$t=0$ (of degree $\kappa(l+1) -1$) and $t=1$ (of degree $l$). Any
meromorphic map can be written as the ratio of two poylnomials,
\begin{equation}
z = \frac{f_1(t)}{f_2(t)},
\end{equation}
where $f_1$ must be of degree $\kappa(l+1)$ and $f_2$ must be of
degree $\kappa l$, from the number of sheets and the behavior at
infinity. Furthermore, near the origin we know that $f_1$ must behave
as $t^{\kappa(l+1)}$ and $f_2$ a constant. Thus, we already have
determined that
\begin{equation}
f_1 \propto t^{\kappa(l+1)}.
\end{equation}
From familiarity with the two-point function map in~\cite{lm1} and
consideration of the above requirements, one can immediately write
down the correct map:
\begin{equation}\label{eq:themap}
z = z_2 \frac{t^{\kappa(l+1)}}{\big[t^{l+1}-(t-1)^{l+1}\big]^\kappa}.
\end{equation}
Note that when $\kappa=1$, the map reduces to the usual two-point map
between $0$ and $z_2$; and when $l=0$, the map reduces to
\begin{equation}
z = z_2 t^\kappa,
\end{equation}
which is the correct map for a two-point function between $0$ and
infinity. Furthermore, that the map is simply raising the two-point
function map to the $\kappa$ power makes sense, since when one goes to
the $\kappa$-cover the correlator should become just the order $(l+1)$
two-point function as depicted in Figure~\ref{fig:branch-cuts}. The
derivative of the map is
\begin{equation}
\der{z}{t} = \frac{\kappa(l+1) z_2}{\big[t^{l+1} - (t-1)^{l+1}\big]^{\kappa+1}}t^{\kappa(l+1)-1}
	(t-1)^l,
\end{equation}
which has zeroes at the correct points. One can see from the map that
it only has $l+1$ distinct images of infinity, as required.

The behavior of the map near the irregular points is needed for the
computation. Near the origin
\begin{equation}
z \approx (-1)^{\kappa l} z_2 t^{\kappa (l+1)}\qquad 
\der{z}{t}\approx (-1)^{\kappa l}\kappa(l+1) z_2t^{\kappa(l+1) -1}.
\end{equation}
Near $t=1$,
\begin{equation}
z\approx z_2 + z_2 \kappa (t-1)^{l+1}\qquad 
\der{z}{t}\approx z_2 \kappa(l+1) (t-1)^l.
\end{equation}
For large $t$,
\begin{equation}
z \approx \frac{z_2}{(l+1)^\kappa}t^\kappa\qquad
\der{z}{t} \approx \frac{z_2\kappa}{(l+1)^\kappa} t^{\kappa -1}.
\end{equation}

The images of infinity at finite $t$ are the same as for the $l+1$
order two-point map, that is
\begin{equation}
t_\infty^{(j)} = \frac{1}{1-\alpha_j}\qquad (\alpha_j)^{l+1}=1,
\end{equation}
where the $\alpha_j$ are the $(l+1)$ roots of unity:
\begin{equation}
\alpha_j = e^{2\pi i\frac{j}{l+1}}.
\end{equation} 
Note that $\alpha_0=1$, which corresponds to $t\to\infty$; the case we
already covered. Therefore, the images of infinity at finite $t$
correspond to $j=1,\dots, l$. If we let
\begin{equation}
t = \frac{1}{1-\alpha_j} + x,
\end{equation}
for small $x\in\co$, then near the images of infinity the map behaves
as
\begin{equation}
z \approx (-1)^\kappa z_2 \left[\frac{\alpha_j}{(l+1)(1-\alpha_j)^2} \frac{1}{x}\right]^\kappa.
\end{equation}
Plugging back in with $t$, one finds the behavior of the map near the
finite images of infinity:
\begin{equation}\label{eq:behavior-finite-images}
\begin{split}
z &\approx (-1)^\kappa
	z_2\left[\frac{\alpha_j}{(l+1)(1-\alpha_j)^2}\right]^\kappa
	\frac{1}{(t-t_\infty^{(j)})^\kappa}\\
\der{z}{t} &\approx (-1)^{\kappa+1}
	\kappa z_2 \left[\frac{\alpha_j}{(l+1)(1-\alpha_j)^2}\right]^\kappa
	\frac{1}{(t-t_\infty^{(j)})^{\kappa+1}}.
\end{split}
\end{equation}
The map is regular at all other points not considered above.

In Appendix~\ref{sec:liouville-action}, we use the local behavior of
the map given above to compute the various contributions to the
Liouville action.

\subsection{Normalization}\label{sec:norm}

The regularized, but unnormalized three-point correlator is given by
\begin{equation}\label{eq:corr-def}
\vev{\underbrace{\sigma^{\tilde{\veps}}_\kappa\cdots\sigma^{\tilde{\veps}}_\kappa}_{l+1}(\infty)\,
	\sigma^\veps_{l+1}(z_2)\,\sigma^\veps_{\kappa(l+1)}(0)}_\delta
 = e^{S_L}\frac{Z^{(\hat{s})}_{\delta'}}{(Z_\delta)^{\kappa(l+1)}}
 = e^{S_L}\,Q^{1-\kappa(l+1)}{\delta'}^{-\frac{c}{3}}\delta^{\frac{c}{3}\kappa(l+1)},
\end{equation}
where for a sphere
\begin{equation}
Z_\delta = Q\delta^{-\frac{c}{3}}\qquad
Z^{(\hat{s})}_{\delta'} = Q{\delta'}^{-\frac{c}{3}}.
\end{equation}
The factor $Q$ is the path integral with the $\delta$ or $\delta'$
dependence, respectively, extracted. The fact that it is the same $Q$
in both expressions is true only because we chose the fiducial
covering space metric to be the same as the base space metric. One can
see that the the path integral scales in this way from the curvature
term of the Liouville action~\cite{lm1}.

Note that instead of inserting the final state twists at a point $z_3$
in the finite $z$-plane and then taking a limit as in
Equation~\eqref{eq:three-point-function}, we have inserted the
operators directly at ``infinity'' (the center of the second half of
the $t$-sphere). As we shall see, the regularizing parameter
$1/\delta$ plays the role of $z_3$.

Calculations in Appendix~\ref{sec:liouville-action} culminate in an
expression for the total Liouville action,
\begin{equation}\begin{split}
S_L^{\text{tot.}} = -\frac{c}{12}\biggr\{&\frac{(\kappa+1)l(l+2)}{\kappa(l+1)}\log|z_2|\\
                              &-\frac{(l+2)^2}{l+1}\log\kappa
                               -4\log(l+1)\\
                              &+\frac{1}{l+1}\left(l^2(\kappa+1)+2l(\kappa -1) 
		+ \frac{(\kappa-1)^2}{\kappa}\right)\log\veps\\
                              &+\frac{(l+1)(\kappa-1)^2}{\kappa}\log\tilde{\veps}\\
                              &+\frac{2(l+1)(\kappa^2+1)}{\kappa}\log\delta\\
                              &-4\log\delta'\biggr\}.
\end{split}
\end{equation}
Note that the $\delta'$ dependence of $S_L$ cancels out with the
$\delta'$ dependence in Equation~\eqref{eq:corr-def}, as it should.
The total $\delta$-dependence in Equation~\eqref{eq:corr-def} is given
by the power
\begin{equation}
\delta:\qquad \frac{c}{6}\frac{(l+1)(\kappa^2-1)}{\kappa} =  4 (l+1)\Delta_\kappa.
\end{equation}
This power of $\delta$ gets cancelled out when demanding that the
insertion of the twist operators at infinity make a properly
normalized bra, as in Equation~\eqref{eq:three-point-function}.

We want to compute the correlator for twist operators that are
normalized such that their two-point function gives unity when they
are unit separated in the $z$-plane. Normalizing the twists in the
finite $z$-plane, we use the result from~\cite{lm1}:
\begin{equation}\label{eq:normalized-twists}
\sigma_n = \frac{1}{\sqrt{C_n\veps^{A_n}Q^{B_n}}}\sigma_n^\veps = D_n\sigma_n^\veps,
\end{equation}
where
\begin{equation}
A_n = -\frac{c}{6}\frac{(n-1)^2}{n}\qquad
B_n = 1-n\qquad
C_n = n^\frac{c}{3}.
\end{equation}

We normalize the twists at infinity in the same way, replacing $\veps$
by $\tilde{\veps}$: at unit separationg in their local coordinates
they should have unit correlator with themselves. From
Appendix~\ref{ap:two-point}, we see that if we normalize the twists at
infinity in this way, then
\begin{equation}
\vev{\sigma_\kappa(\infty)\sigma_\kappa(0)}_{\delta} = \delta^{4\Delta_\kappa}.
\end{equation}
To ensure that the final state corresponds to a normalized bra, then,
we should use
\begin{equation}
\lim_{z_3\to\infty}|z_3|^{4\Delta_\kappa}\sigma_\kappa(z_3) 
  = \delta^{-4\Delta_\kappa}\sigma_\kappa(\infty).
\end{equation}
Thus, the $\delta$ dependence cancels out.  Furthermore, once we
normalize the twist operators the $\veps$ dependence drops out:
\begin{equation}
\veps:\qquad  -\frac{c}{12}\frac{1}{l+1}\left(l^2(\kappa+1)+2l(\kappa -1) 
		+ \frac{(\kappa-1)^2}{\kappa}\right)
	-\frac{1}{2}(A_{l+1} +A_{\kappa(l+1)}) = 0.
\end{equation}
The $\tilde{\veps}$ dependence also cancels out. Now, let's check the
powers of $Q$:
\begin{equation}
Q:\qquad 1-\kappa(l+1) + \frac{l}{2} + \frac{\kappa(l+1)-1}{2}
		+ \frac{(\kappa-1)(l+1)}{2} = 0;
\end{equation}
there is no $Q$-dependence left and so we never have to actually
compute the path integral. One can show that the $Q$-dependence always
cancels out for correlators with spherical covering spaces.

One finds that the powers of $(l+1)$ cancel out as well. The constant
$C$, then, is given as a power of $\kappa$:
\begin{equation}
|C|^2 = \kappa^{-\frac{c}{12}\frac{l(l+2)}{l+1}}.
\end{equation}
The power of $\kappa$ may seem mysterious, until one realizes that
\[
\Delta_{l+1} = \frac{c}{24}\left(l+1 - \frac{1}{l+1}\right)
	= \frac{1}{4}\frac{l(l+2)}{l+1},
\]
and so we write
\begin{equation}
|C|^2 = \kappa^{-2\Delta_{l+1}}.
\end{equation}

\subsection{The twist jacobian factor $T$}

Plugging in the value of $C$ into Equation~\eqref{eq:C-gives-T}, one
finds the twist Jacobian factor,
\begin{equation}
T = \kappa^{-2\Delta_{l+1}}\frac{|u_2|^{4\Delta_{l+1}}}
	{|z_2|^{2\frac{\kappa+1}{\kappa}\Delta_{l+1}}}.
\end{equation}
Note that this Jacobian factor cannot naturally be thought of as the
product of Jacobian factors from the different twist operators. This
fact is what makes correlation functions of twist operators
nontrivial.

\section{Computing the mode jacobian factor $M$}\label{sec:M}

To compute the mode Jacobian factor, $M$, we first illustrate how
modes transform under the map $z=bu^\kappa$. Then, we compute the mode
Jacobian factor from the initial state modes, followed by the mode
Jacobian factor from the vertex operator modes. The final state does
not have any modes, and therefore does not contribute to $M$. In
discussing the initial state, we find that we must also consider the
$\kappa$-dependence of the normalization of the state.

\subsection{Transformation law for modes}

First, let us examine how the different modes transform under the map,
\begin{equation}
z = b u^\kappa.
\end{equation}
Modes that circle the origin get opened up by a factor of $\kappa$,
and also get a Jacobian factor. They transform as
\begin{equation}\label{eq:mode-trans}
\mathcal{O}_m^{(z)} = b^m \kappa^{1-\Delta}\mathcal{O}^{(u)}_{m\kappa},
\end{equation}
where $\Delta$ is the weight of the \emph{field} $\mathcal{O}(z)$. One
can see this from the definition of modes in the $\kappa(l+1)$-twisted
sector~\cite{lm2}:
\begin{equation}
\mathcal{O}_{\frac{m}{\kappa(l+1)}}^{(z)} = \oint\frac{\drm z}{2\pi i}\sum_{j=1}^{\kappa(l+1)}
\mathcal{O}_{(j)}(z)e^{2\pi i\frac{m}{\kappa(l+1)}(j-1)}z^{\Delta - 1 + \frac{m}{\kappa(l+1)}}.
\end{equation}
The subscripted parenthetical index on the field $\mathcal{O}_{(j)}$
denotes the copy the field lives on.

Modes which circle $z_2$, where there is no branching point, transform
as
\begin{equation}
\mathcal{O}^{(z)}_m = \left(\der{z}{u}\bigg|_{u_2,z_2}\right)^m\mathcal{O}_m^{(u)}.
\end{equation}
Near the point $u_2$, the map behaves as
\begin{calc}
z &\approx z_2 + \kappa b u_2^{\kappa-1}(u-u_2) \\
  &= z_2 + \kappa \frac{z_2}{u_2}(u-u_2).
\end{calc}
Therefore, we see that, under this map, modes which circle the point
$z_2$ transform as
\begin{equation}
\mathcal{O}^{(z)}_{m} = \left(\kappa \frac{z_2}{u_2}\right)^m\mathcal{O}_m^{(u)}.
\end{equation}

Since the spectral flowed final state does not have any modes, we can
write $M$ as a product of Jacobian factors from the initial state and
Jacobian factors from the vertex operator,
\begin{equation}
M = M_i M_v.
\end{equation}

\subsection{The mode jacobian factor from the initial state}\label{sec:Mi}

For the initial state, we need to be a little careful about the
normalization. Consider an initial state defined in the base space as
\begin{equation}\label{eq:state-base}
\ket{\psi^{(z)}} = \mathcal{C}\mathcal{O}_m^{(z)}\ket{\sigma_{\kappa(l+1)}}.
\end{equation}
The norm of this state is given by
\begin{calc}
\braket{\psi^{(z)}|\psi^{(z)}} 
	&= |\mathcal{C}|^2\bra{\sigma_{\kappa(l+1)}}
		\mathcal{O}^{\dg(z)}_{-m}\mathcal{O}^{(z)}_m\ket{\sigma_{\kappa(l+1)}}\\
	&= |\mathcal{C}|^2\bra{\sigma_{l+1}}
		\left(b^{-m}\kappa^{1-\Delta}\mathcal{O}^{\dg(u)}_{-m\kappa}\right)
		\left(b^m\kappa^{1-\Delta}\mathcal{O}^{(u)}_{m\kappa}\right)\ket{\sigma_{l+1}}\\
	&= |\mathcal{C}|^2\kappa^{2(1-\Delta)}\bra{\sigma_{l+1}}\mathcal{O}^{\dg(u)}_{-m\kappa}
		\mathcal{O}^{(u)}_{m\kappa}\ket{\sigma_{l+1}}.
\end{calc}

On the other hand, suppose one were to define the state directly in
the cover as
\begin{equation}
\ket{\psi^{(u)}} = \mathcal{C}_\text{cover}\mathcal{O}_{m\kappa}^{(u)}\ket{\sigma_{l+1}},
\end{equation}
then its norm is given by
\begin{equation}
\braket{\psi^{(u)}|\psi^{(u)}} = |\mathcal{C}_\text{cover}|^2
\bra{\sigma_{l+1}}\mathcal{O}^{\dg(u)}_{-m\kappa}\mathcal{O}^{(u)}_{m\kappa}\ket{\sigma_{l+1}}.
\end{equation}
Requiring that both $\ket{\psi^{(z)}}$ and $\ket{\psi^{(u)}}$ be
normalized, means
\begin{equation}
\mathcal{C} = \frac{\mathcal{C}_\text{cover}}{\kappa^{1-\Delta}}.
\end{equation}
Thus, using Equation~\eqref{eq:mode-trans} with
Equation~\eqref{eq:state-base}, one finds
\begin{equation}
\ket{\psi^{(z)}} = b^m\ket{\psi^{(u)}}.
\end{equation}
From this we see that the $\kappa$ factor in
Equation~\eqref{eq:mode-trans} gets cancelled out by the $\kappa$s one
gets in normalizing the state in the covering space. This argument
easily generalizes to all of the modes acting in the initial state.

The only contributions to $M_i$, therefore, come from the factors
$b^m$. Because of the way they come in, we see that the the total
factor we get (from the left) is
\begin{equation}
b^{-h^\text{modes}_i},
\end{equation}
where $h^\text{modes}_i$ is the weight of all of the modes in the
initial state. This is simply the total weight of the initial state
minus the weight of the bare twist:
\begin{calc}
h^\text{modes}_i &= h'_{i} - \Delta_{\kappa(l+1)}\\
	&= \frac{l+1}{4}\left(\cancel{\kappa}-\frac{1}{\kappa}\right)
		+\frac{1}{\kappa}\left(\frac{l}{2} + 1\right) + \frac{N}{\kappa}
		-\frac{1}{4}\left(\cancel{\kappa(l+1)} - \frac{1}{\kappa(l+1)}\right)\\
	&= \frac{1}{\kappa}\left[
	\frac{l}{2} +N + 1 - \Delta_{l+1}
		\right].
\end{calc}

Therefore, the total contribution (including both the left and right
sectors) to the mode Jacobian factor from the initial state is given
by
\begin{calc}
M_i &= |b|^{-\frac{2}{\kappa}\left[\frac{l}{2} + 1 - \Delta_{l+1}\right]}
	b^{-\frac{N}{\kappa}}\bar{b}^{-\frac{\bar{N}}{\kappa}}\\
     &= \frac{|u_2|^{2\left[\frac{l}{2} + 1 - \Delta_{l+1}\right]}}
	{|z_2|^{\frac{2}{\kappa}\left[\frac{l}{2} + 1 - \Delta_{l+1}\right]}}
        \frac{u_2^N\bar{u}_2^{\bar{N}}}{z_2^\frac{N}{\kappa}\bar{z}_2^\frac{\bar{N}}{\kappa}}.
\end{calc}

\subsection{The mode jacobian factor from the vertex operator}

The vertex operator, we can see from
\[
\mathcal{O}^{(z)}_{m} = \left(\kappa \frac{z_2}{u_2}\right)^m\mathcal{O}_m^{(t)},
\]
contributes a factor of
\begin{equation}
M_v = \left(\kappa\tfrac{z_2}{u_2}\right)^{-h^\text{modes}_v}
      \left(\kappa\tfrac{\bar{z}_2}{\bar{u}_2}\right)^{-\bar{h}^\text{modes}_v},
\end{equation}
where $h^\text{modes}_v$ is the weight of the vertex operator less the
weight of the bare $(l+1)$-twist, that is,
\begin{equation}
h^\text{modes}_v = \bar{h}^\text{modes}_v
 = \frac{l}{2} + 1 - \Delta_{l+1}.
\end{equation}
Therefore, we see that the contribution to the mode Jacobian factor
from the vertex operator is given by
\begin{equation}
M_v = \kappa^{2\Delta_{l+1} - (l+ 2)}\left(\frac{|u_2|}{|z_2|}\right)^{l+2 - 2\Delta_{l+1}}.
\end{equation}

\section{The CFT amplitude}\label{sec:A-prime}

Multiplying the mode Jacobian factor,
\begin{equation}
M = M_iM_v = \kappa^{2\Delta_{l+1} - (l+2)}\frac{|u_2|^{2(l+2) - 4\Delta_{l+1}}}
	{|z_2|^{\frac{\kappa+1}{\kappa}[l+2 - 2\Delta_{l+1}]}}
	\frac{u_2^N\bar{u}_2^{\bar{N}}}{z_2^\frac{N}{\kappa}\bar{z}_2^\frac{\bar{N}}{\kappa}},
\end{equation}
with the twist Jacobian factor,
\[
T = \kappa^{-2\Delta_{l+1}}\frac{|u_2|^{4\Delta_{l+1}}}
	{|z_2|^{2\frac{\kappa+1}{\kappa}\Delta_{l+1}}},
\]
one finds
\begin{equation}
TM = \kappa^{-(l+2)}\frac{|u_2|^{2(l+2)}}{|z_2|^{\frac{\kappa+1}{\kappa}(l+2)}}
     \frac{u_2^N\bar{u}_2^{\bar{N}}}{z_2^\frac{N}{\kappa}\bar{z}_2^\frac{\bar{N}}{\kappa}}.
\end{equation}

Finally, from~\cite{acm1} we take the $\kappa=1$ CFT amplitude (before
spectral flow and \emph{without Jacobian prefactor $|z|^{l+2}$ from
mapping from the cylinder to the complex plane})
\begin{equation}\label{eq:kappa1amp}
{\mathcal{A}'}^{\kappa=1} = (-1)^{k+\bar{k}}
	\sqrt{\choose{N+l+1}{N}\choose{\bar{N} + l+1}{\bar{N}}}
	\frac{1}{|u_2|^{2(l+2)}u_2^N\bar{u}_2^{\bar{N}}},
\end{equation}
and use the Jacobian factor $TM$ to find
\begin{equation}\label{eq:final-primed-amp}
{\mathcal{A}'}^\kappa(z_2, \bar{z}_2) = \kappa^{-(l+2)}
	\sqrt{\choose{N+l+1}{N}\choose{\bar{N} + l+1}{\bar{N}}}
	\frac{1}{|z_2|^{\frac{\kappa+1}{\kappa}(l+2)}
		z_2^\frac{N}{\kappa}\bar{z}_2^\frac{\bar{N}}{\kappa}},
\end{equation}
which reduces to Equation~\eqref{eq:kappa1amp} for $\kappa=1$. 


\section{Combinatorics}\label{sec:comb}

As in~\cite{acm1}, we want to consider an initial state with $\nu$
excitations that de-excites into a final states with $\nu-1$
excitations (shown in Figure~\ref{fig:combinatorics-states}); however,
the background geometry now consists of $\kappa$-twisted component
strings. We assume, therefore, that $\kappa$ divides $N_1N_5$. Because
the theory is orbifolded by $S_{N_1N_5}$, the initial state, the final
state, and the vertex operator must all be symmetrized over the
$N_1N_5$ copies.

We relate the full symmetrized states and operator to the amplitude
computed above, in which we do not worry about these combinatoric
factors.  The combinatorics for this problem are quite similar to and
for $\kappa=1$ reduce to the combinatorics in~\cite{acm1}.

\subsection{The initial state}

We write the initial state as a sum over all the symmetric
permutations:
\begin{equation}\label{eq:sym-nu-sum}
\ket{\Psi_\nu} = \mathcal{C}_\nu\bigg[\ket{\psi_\nu^1} + \ket{\psi_\nu^2} + \dots\bigg],
\end{equation}
where $\mathcal{C}_\nu$ is the overall normalization and each
$\ket{\psi^i_\nu}$ is individually normalized. The initial state for
$\nu=2$, $\kappa=2$, and $l=2$ is shown in
Figure~\ref{fig:unphy-comb-in}.

To understand what we are doing better, note that the state
$\ket{\psi_{\nu=2}^1}$ with $\kappa =2$ and $l+1=3$ can be written
schematically as
\begin{equation}\label{eq:schematic-nu-state}
\ket{\psi_\nu^1} = \Ket{
	\big[1 \cdots 6\big]
	\big[7 \cdots 12\big]
	\big[12,13]\big[14,15]\cdots}
\end{equation}
where the numbers in the square brackets are indicating particular ways
of twisting individual strands corresponding to particular cycles of
the permutation group. For instance,
\begin{equation}
\ket{[1234]},
\end{equation}
indicates that we twist strand 1 into strand 2 into strand 3 into
strand 4 into strand 1 and leave strands 5 through $N_1N_5$ untwisted.
In Equation~\eqref{eq:schematic-nu-state}, the first two $6$-twists
are the two $\kappa(l+1)$-twists which indicate the two
excitations; the remaining $(\kappa=2)$-twists correspond to the
background.

To determine the normalization $\mathcal{C}_\nu$, we need to count the
number of distinct terms in Equation~\eqref{eq:sym-nu-sum}. To count
the number of states, we imagine constructing one of the terms and
count how many choices we have. First, of the $N_1N_5$ strands we must
pick $\nu\kappa(l+1)$ of them to make into the $\nu$ excited states.
Next, we must break the $\nu\kappa(l+1)$ strands into sets of
$\kappa(l+1)$ to be twisted together. Then, we must pick a way of
twisting together each set of $\kappa(l+1)$. Finally, we must make
similar choices for the $N_1N_5 - \nu\kappa(l+1)$ strands that are
broken into the $\kappa$-twists of the background. Putting these
combinatoric factors together, one finds
\begin{calc}
N_\text{terms} &= \choose{N_1N_5}{\nu\kappa(l+1)}\\
 &\qquad \times \choose{\nu\kappa(l+1)}{\kappa(l+1)}\choose{(\nu-1)\kappa(l+1)}{\kappa(l+1)}
                   \cdots \choose{\kappa(l+1)}{\kappa(l+1)}\frac{1}{\nu!}\times
                   \big([\kappa(l+1)-1]!\big)^\nu\\
 &\qquad \times \choose{N_1N_5 - \nu\kappa(l+1)}{\kappa}
                \cdots \choose{\kappa}{\kappa}
                \frac{1}{\left[\frac{N_1N_5}{\kappa} - \nu(l+1)\right]!}
                \times \big((\kappa -1)!\big)^{\frac{N_1N_5}{\kappa} - \nu(l+1)}\\
  &= \frac{(N_1N_5)!}{[\kappa(l+1)]^\nu\, \kappa^{\frac{N_1N_5}{\kappa} - \nu(l+1)}\,
                   \left(\frac{N_1N_5}{\kappa} - \nu(l+1)\right)!\,\nu!}.
\end{calc}
Choosing $\mathcal{C}_\nu$ to be real, one finds
\begin{equation}
\mathcal{C}_\nu = \frac{1}{\sqrt{N_\text{terms}}}
                = \left[
                  \frac{(N_1N_5)!}{[\kappa(l+1)]^\nu\, \kappa^{\frac{N_1N_5}{\kappa} - \nu(l+1)}\,
                   \left(\frac{N_1N_5}{\kappa} - \nu(l+1)\right)!\,\nu!}
                  \right]^{-\frac{1}{2}}.
\end{equation}

\subsection{The final state}

\begin{figure}[t]
\begin{center}
\subfigure[~The initial state with $\nu=2$ excitations.]
{\label{fig:unphy-comb-in}
\includegraphics[width=6cm]{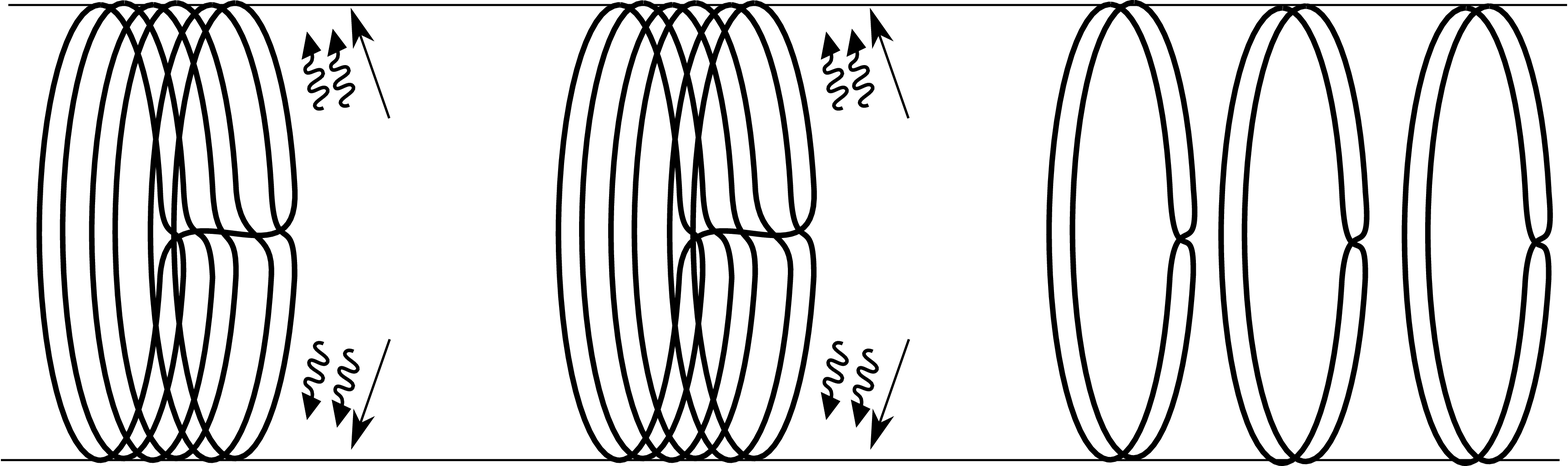}}
\hspace{30pt}
\subfigure[~The final state with $\nu-1=1$ excitations.]
{\label{fig:unphy-comb-fin}
\includegraphics[width=6cm]{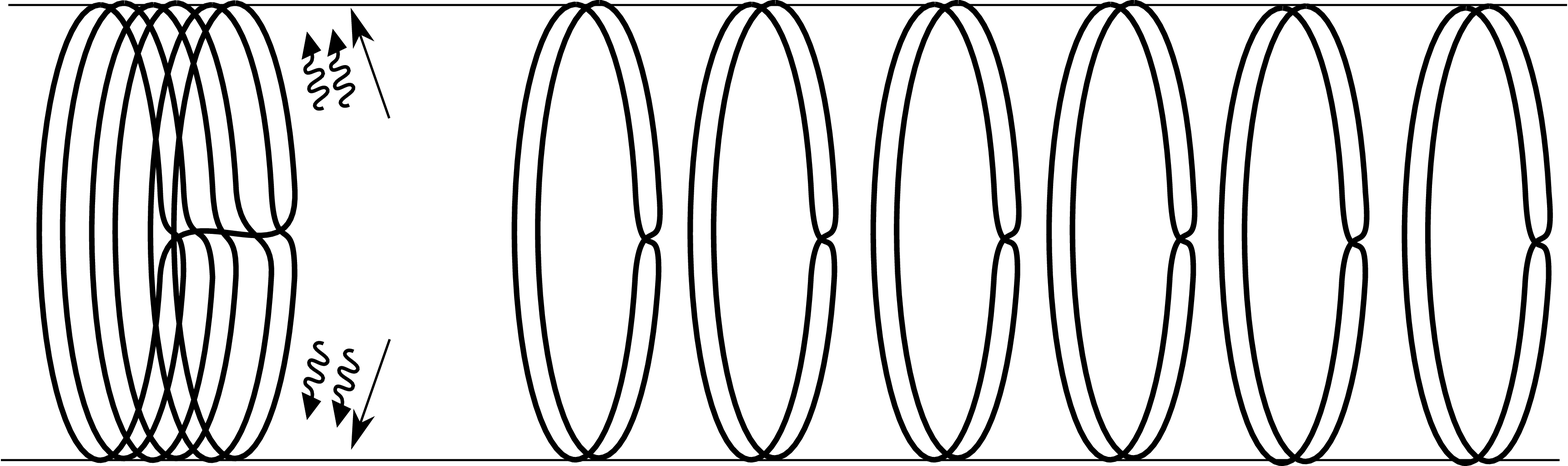}}
\caption{The initial and final states of the unphysical amplitude 
with some excitations present in the background. An $l=2$ transition
from $\nu=2$ excitations to $1$ excitation of the $\kappa=2$
background is shown.}\label{fig:combinatorics-states}
\end{center}
\end{figure}

The final state is simply $\ket{\Psi_{\nu-1}}$ with its corresponding
normalization $\mathcal{C}_{\nu-1}$. This state is shown for $\nu=2$,
$\kappa=2$, and $l=2$ in Figure~\ref{fig:unphy-comb-fin}.

\subsection{The vertex operator}

The normalization of the vertex operator was determined
in~\cite{acm1}. The full symmetrized vertex operator is written
\begin{equation}
\mathcal{V}_\text{sym} = \mathcal{C}\sum_i\mathcal{V}_i,
\end{equation}
where
\begin{equation}
\mathcal{C}=\left[\frac{(N_1N_5)!}{(l+1)[N_1N_5 - (l+1)]!}\right]^{-\frac{1}{2}}.
\end{equation}

\subsection{The amplitude}

The full amplitude of interest is
\begin{equation}
\bra{\Psi_{\nu-1}}\mathcal{V}_\text{sym}\ket{\Psi_\nu},
\end{equation}
which we wish to relate to a ``single'' unsymmetrized amplitude.
Having the normalization of the initial state, the final state, and
the vertex operator, all that remains is to determine which terms in
the symmetrized states and operator combine to give a nonzero
amplitude. For each initial state there are $\nu$ excitations that a
term in the vertex operator can de-excite. We are left with the
question, how many $(l+1)$-twists can untwist a given
$\kappa(l+1)$-twist into $l+1$ $\kappa$-twists?  In fact, there are
$\kappa$ such twist operators, which brings the number of nonzero
amplitudes to
\begin{equation}\label{eq:number-amplitudes}
N_\text{terms}\, \nu\kappa  = \kappa\frac{\nu}{\mathcal{C}_\nu^2}.
\end{equation}

To illustrate, consider the initial state in
Equation~\eqref{eq:schematic-nu-state} for $\nu=2$, $\kappa=2$, and
$l+1=3$. First, there are two different excitations to untwist.
Consider untwisting the first excitation $[123456]$. There are exactly
two $3$-twist vertex operators that untwist the excitation into three
sets of $2$-twists: $[531]$ and $[642]$. The $[531]$ vertex operator
breaks the initial state into $[12][34][56]$, while the $[642]$ vertex
operator breaks the initial state into $[61][23][45]$.

Using Equation~\eqref{eq:number-amplitudes} and the normalizations, we
can relate the total symmetrized amplitude to the amplitude computed
with only one term from the initial state, the final state, and the
vertex operator:
\begin{calc}\label{eq:full-comb-factor}
\bra{\Psi_{\nu-1}} \mathcal{V}_\text{sym} \ket{\Psi_{\nu}}
  &= \mathcal{C}_{\nu-1}\mathcal{C}\mathcal{C}_\nu\, \kappa\frac{\nu}{\mathcal{C}_\nu^2}
     \bra{\psi^1_{\nu-1}}\mathcal{V}_1 \ket{\psi^1_{\nu}}\\
  &= \sqrt{\kappa^{l+2}\nu
            \frac{\left[\frac{N_1N_5}{\kappa}-(\nu-1)(l+1)\right]!}
                 {\left[\frac{N_1N_5}{\kappa}-\nu(l+1)\right]!}
            \frac{\left[N_1N_5 - (l+1)\right]!}{(N_1N_5)!}}     
\bra{\psi^1_{\nu-1}}\mathcal{V}_1 \ket{\psi^1_{\nu}}.
\end{calc}

\subsection{The large $N_1N_5$ limit}

While the expression in Equation~\eqref{eq:full-comb-factor} is
complicated, it simplifies considerably in the large $N_1N_5$ limit of
ultimate interest. We take both $N_1N_5$ and $N_1N_5/\kappa$ to be
large,
\begin{equation}\label{eq:comb-factors-limit}
\frac{\left[\frac{N_1N_5}{\kappa}-(\nu-1)(l+1)\right]!}
                 {\left[\frac{N_1N_5}{\kappa}-\nu(l+1)\right]!}
                \longrightarrow \left(\frac{N_1N_5}{\kappa}\right)^{(l+1)}
\hspace{20pt}
\frac{[N_1N_5-(l+1)]!}{(N_1N_5)!}
	\longrightarrow
	(N_1N_5)^{-(l+1)},
\end{equation}
in which case Equation~\eqref{eq:full-comb-factor} reduces to
\begin{equation}\label{eq:large-N-comb}
\bra{\Psi_{\nu-1}} \mathcal{V}_\text{sym} \ket{\Psi_{\nu}}
  = \sqrt{\kappa\nu}\, \bra{\psi^1_{\nu-1}}\mathcal{V}_1 \ket{\psi^1_{\nu}}.
\end{equation}
The $\sqrt{\nu}$ prefactor can be thought of as a ``Bose enhancement''
effect.

\section{The rate of emission}\label{sec:rate}

We now can put all of the pieces together to find the spectrum and
rate of emission. Plugging Equation~\eqref{eq:final-primed-amp} into
Equation~\eqref{eq:unphysical-to-physical} along with the combinatoric
factors from Equation~\eqref{eq:large-N-comb}, one finds the amplitude
for emission with angular mometum $(l, k+\bar{k}-l, \bar{k}-k)$
\begin{equation}
\mathcal{A}_{l, k+\bar{k}-l, \bar{k}-k} = \sqrt{\nu\kappa}\,\kappa^{-(l+2)}
 	\sqrt{\choose{N+l+1}{N}\choose{\bar{N} + l+1}{\bar{N}}}
	z^{\frac{1}{\kappa}(\frac{l}{2} + N + 1)-\alpha(\frac{l}{2}-k)}
   \bar{z}^{\frac{1}{\kappa}(\frac{l}{2} + \bar{N} + 1) - \bar{\alpha}(\frac{l}{2} - \bar{k})}.
\end{equation}
Plugging back in with the physical cylindrical coordinates,
\begin{equation}
z = e^{\frac{i}{R}(y+t)}\qquad \bar{z} = e^{-\frac{i}{R}(y-t)},
\end{equation}
we can read off the spectrum for emission,
\begin{equation}\begin{split}\label{eq:final-spectrum}
E_0 &= \frac{1}{\kappa R}\left[(\kappa\alpha + \kappa\bar{\alpha} - 2)\tfrac{l}{2}
		- \kappa(\alpha k + \bar{\alpha}\bar{k}) - N - \bar{N} - 2\right]\\
\lambda_0 &= \frac{1}{\kappa R}
	\left[-\kappa(\alpha - \bar{\alpha})\tfrac{l}{2} + \kappa(\alpha k - \bar{\alpha}\bar{k})
	 + N - \bar{N}\right],
\end{split}\end{equation}
where recall that $\alpha = 2n + 1/\kappa$ and
$\bar{\alpha}=2\bar{n}+1/\kappa$.  For there to be emission the energy
of the emitted particle must be positive, meaning that $E_0>0$.

The unit amplitude with the $(\sigma, \tau)$ dependence removed is
\begin{equation}
\mathcal{A}_\text{unit}(0) = \sqrt{\nu} \kappa^{-l -\frac{3}{2}}
 	\sqrt{\choose{N+l+1}{N}\choose{\bar{N} + l+1}{\bar{N}}}.
\end{equation}
Section~\ref{sec:comb}, where $\nu$ is defined, calculates the
combinatorics for the \emph{unphysical} amplitude. For the
\emph{physical} amplitude, the combinatorics are the same except that
the emission process is a transition from $\nu-1$ to $\nu$
$\kappa(l+1)$-twists. Thus, the above is the amplitude for emission of
the $\nu$th particle. Plugging into
Equation~\eqref{eq:D1D5-decay-rate}, one finds the emission rate for
the $\nu$th particle,
\begin{equation}\label{eq:final-rate}
\der{\Gamma}{E} = \nu\kappa^{-2l-3}
	\frac{2\pi}{2^{2l+1}\,l!^2}\frac{(Q_1Q_5)^{l+1}}{R^{2l+3}}(E^2-\lambda^2)^{l+1}\,
	\choose{N+l+1}{N}\choose{\bar{N} + l+1}{\bar{N}}
	\delta_{\lambda,\lambda_0}\delta(E - E_0),
\end{equation}
with energy and momentum in Equation~\eqref{eq:final-spectrum} and
angular momentum ${(l, k+\bar{k}-l,\bar{k}-k)}$. This answer exactly
matches the gravity calculation in~\cite{cm3}.

\section{Discussion}\label{sec:discussion}

We reproduced the full spectrum and emission rate of supergravity
minimal scalars from the geometries found in~\cite{ross} by using a CFT
formalism. In~\cite{cm1,cm2,cm3}, using a heuristic picture of the CFT
process the spectrum and rate was found, but only for special cases
($N=\bar{N}=0$ and $k=\bar{k}=0$) and the normalization of the
heuristic vertex operator was determined only
indirectly. In~\cite{acm1}, the full rate and spectrum was found as a
rigorous calculation for $\kappa=1$ with a vertex operator whose
coupling to flat space and normalization was determined directly from
the AdS/CFT correspondence.

The new feature of this paper from results found in~\cite{acm1} is the
$\kappa$-dependence. Using the old effective string description, the
$\kappa$-dependence in Equations~\eqref{eq:final-spectrum}
and~\eqref{eq:final-rate} could have been guessed via a heuristic
argument that we now describe. Taking higher values of $\kappa$
corresponds to twisting the background strings by $\kappa$. The
process, then, looks the same as for $\kappa=1$ but taking $R \mapsto
\kappa R$. This reproduces the explicit $\kappa$ dependence of
Equation~\eqref{eq:final-rate}, but the spectrum is slightly more
complicated. In the spectrum, we also should take $R
\mapsto \kappa R$, since the energy level spacing becomes reduced by a
factor of $\kappa$; however, things are complicated by the parameters
$n$ and $\bar{n}$. Why does the $n$ and $\bar{n}$ part of the spectrum
get multiplied by $\kappa$ with respect to the rest of the
contributions to the energy? The parameters $n$ and $\bar{n}$ control
the Fermi level of the physical initial state. In the $\kappa$-cover,
the fermions fill up to energy level $\kappa n$ and not $n$; thus, the
extra factor of $\kappa$.

With the final answer and CFT computation in front of us, this
heuristic argument seems compelling indeed; however, we argue that it
is not completely obvious a priori that the effective string reasoning
works for this calculation. Certainly, the way that the
$\kappa$-dependence works out in the rigorous CFT calculation seems
quite nontrivial and nonobvious. That the action of the vertex
operator so simply just gets ``divided by'' $\kappa$ does not seem
obvious. In any case, one of our goals in this paper is to demonstrate
the formalism in~\cite{acm1} and put the reasoning in~\cite{cm1, cm2,
cm3} on a firmer footing. Along the way, we found the form of
supergravity excitations on the orbifolded-AdS background; it would be
nice to better understand the identification of the supergravity
multiplet in orbifolded-AdS.

That the rate of emission computed in this class of geometries can be
reproduced via a CFT calculation may seem insignificant in the face of
so many AdS/CFT successes. There are two reasons why this calculation
is of interest. First, most AdS/CFT calculations use a gravity
calculation in the AdS to compute a CFT correlator, whereas we
use~\cite{acm1} to compute the rate of emission \emph{out of the} CFT
or AdS. This demonstrates the formalism of~\cite{acm1}, which hearkens
back to and puts a firmer footing on the ``effective string''
calculations
of~\cite{stromvafa,radiation-1,radiation-2,radiation-3,radiation-4,
radiation-5}. Second, as explained in~\cite{cm1, cm2, cm3}, the CFT
calculation justifies interpreting the ergoregion emission as Hawking
radiation from these, albeit nongeneric, microstates of a black
hole. This may help us better understand black holes in string theory
and thereby quantum gravity.

All of the calculations in this paper were performed on the ``orbifold
point'' of the D1D5 CFT. The emission process described in the gravity
side should be dual to the CFT off of the orbifold point, and so it is
an open question why this and many other calculations on the orbifold
point so accurately reproduce the gravitational physics. In the
future, we plan to move off of the orbifold point in the hope that we
can answer this and many dynamical questions about black holes.

\section*{Acknowledgments}

We are very grateful for S.~Mathur's insights, guidance, and helpful
discussions over the course of this project. We also thank
Y.~Srivastava and P.~Kraus for conversations and comments.

This work was supported in part by DOE grant~DE-FG02-91ER-40690.

\appendix

\section{Computing the Liouville action}\label{sec:liouville-action}

The correct way to define the correlator in
Equation~\eqref{eq:three-point-function} is to cut holes in the $z$
plane around $z=0$ and $z=z_2$ of size $\veps$ and then a large cut at
$|z|=1/\delta$. The small holes around the origin and $z_2$ have the
appropriate boundary conditions for the twist operators. Upon the cut
at $|z|=1/\delta$, we glue a second flat disc with coordinate
$\tilde{z}$. The center of this second disc has a hole of size
$\tilde{\veps}$, which has the boundary conditions appropriate for the
$(l+1)$ $\kappa$-twists at infinity. The metric on this manifold is
\begin{equation} 
\begin{split}
ds^2 &= \left\{ 
\begin{array}{rl} 
 dzd\bar{z} & \qquad |z| <\frac{1}{\delta}\\ 
 \\
 d\tilde{z}d\bar{\tilde{z}} & \qquad |\tilde{z}| < \frac{1}{\delta}
\end{array} \right. \\
z &= \frac{1}{\delta^2\tilde{z}}.
\end{split} \label{Eqn:BaseMetric}
\end{equation} 

\begin{figure}[ht]
\begin{center}
\subfigure[~The regularized $z$-sphere.]{
\includegraphics[width=6cm]{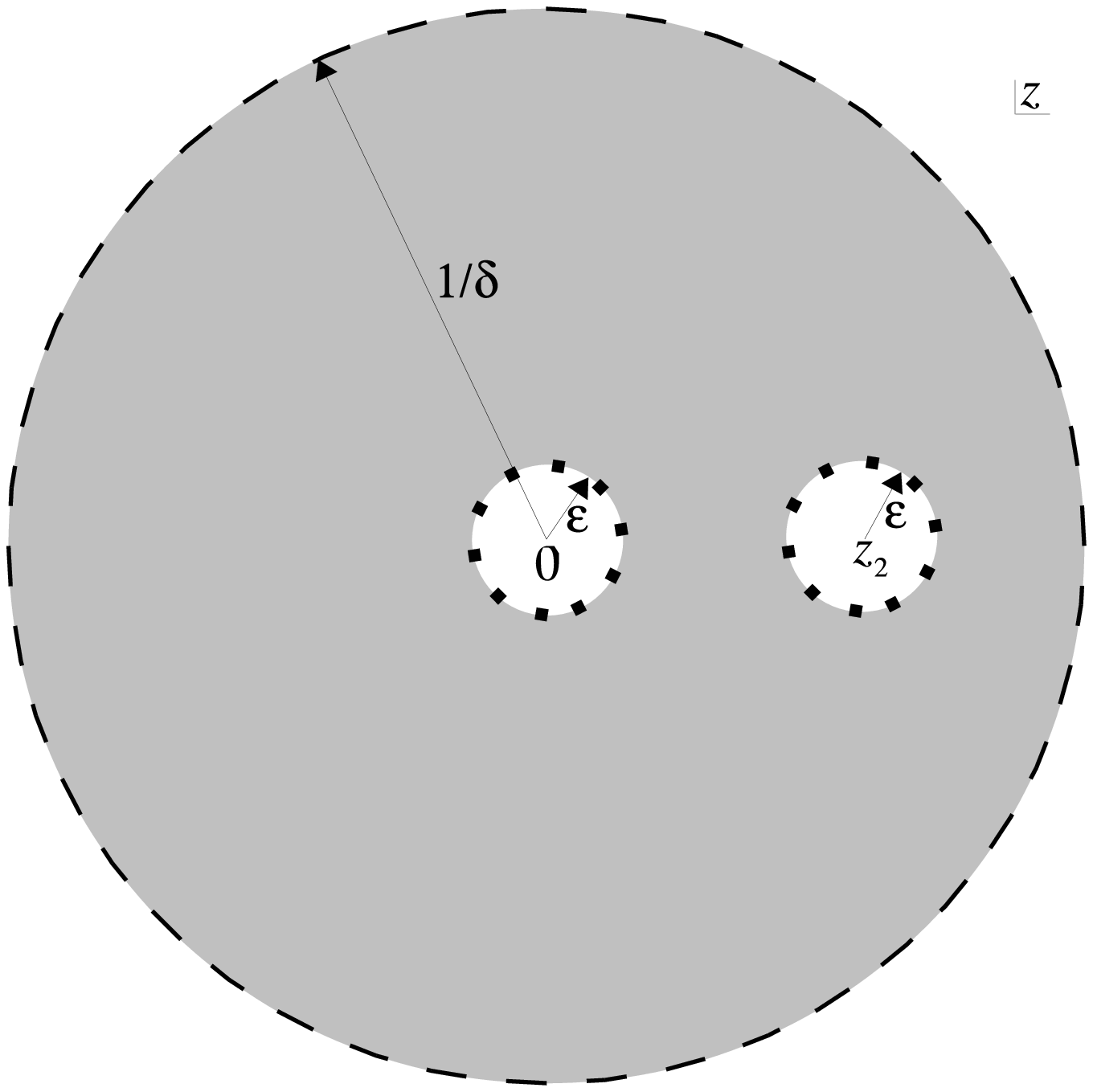}
\includegraphics[width=6cm]{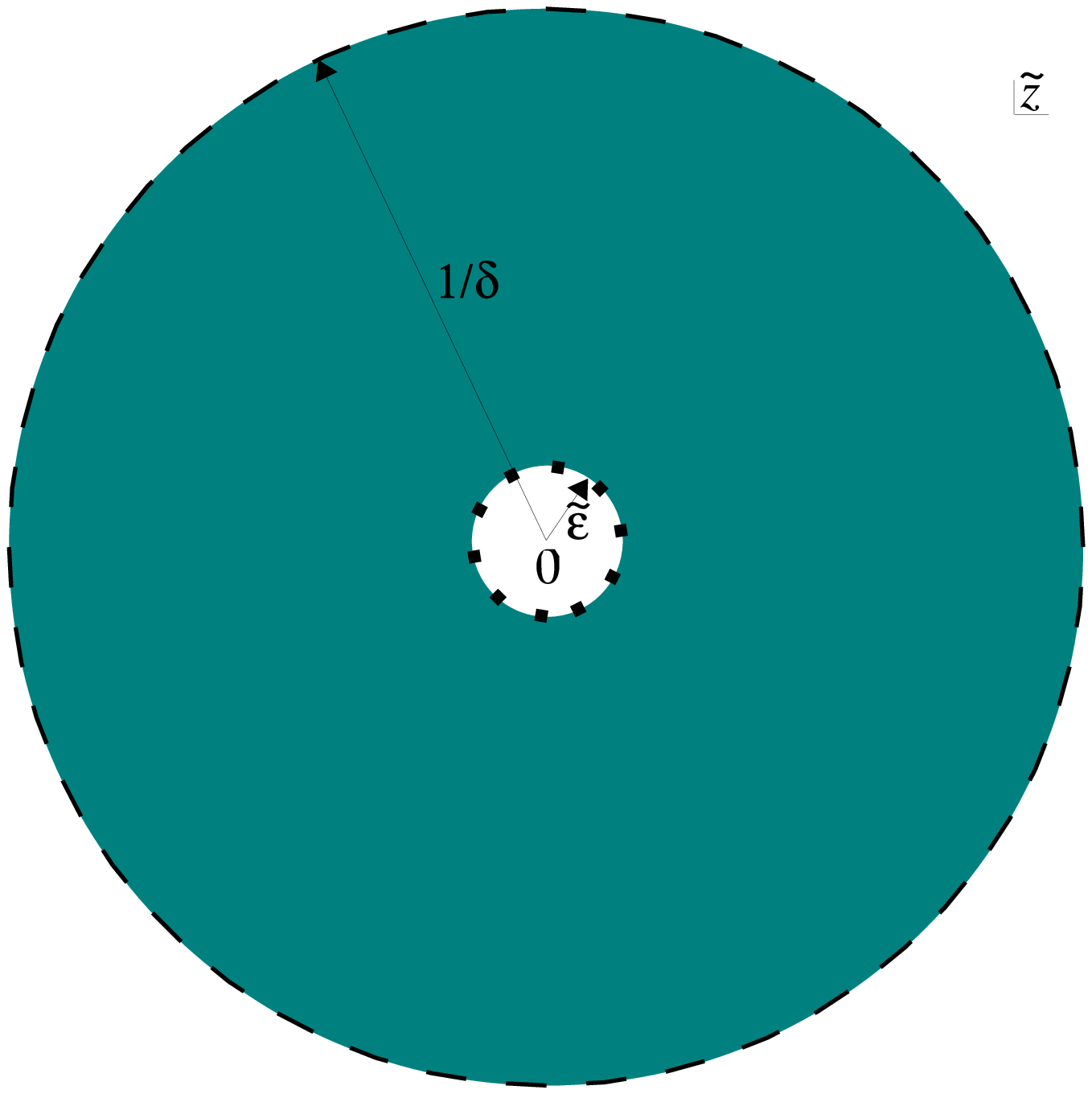}}\\
\vspace{10pt}
\subfigure[~The $t$-sphere for $l=3$.]{
\includegraphics[width=6cm]{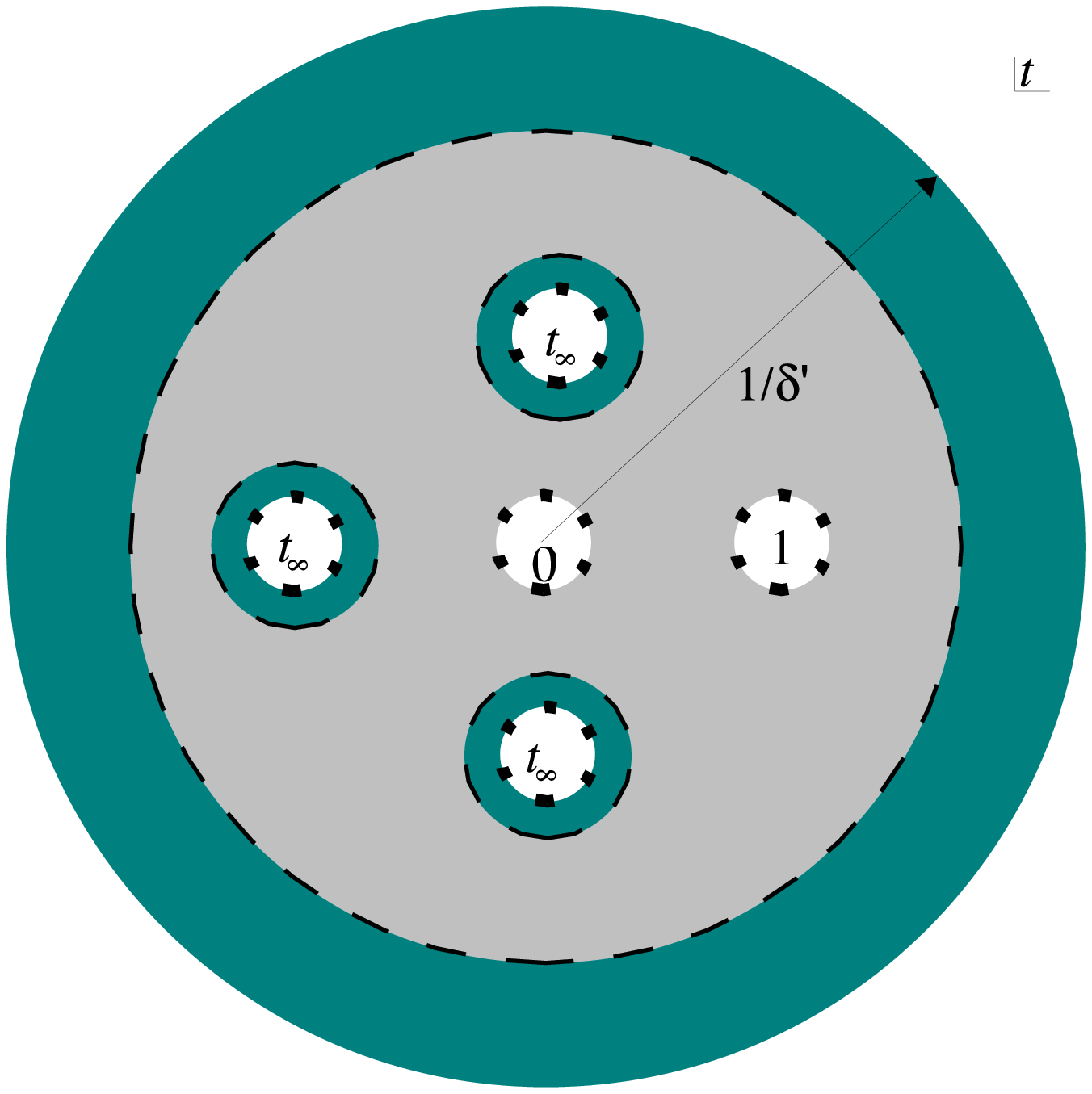}
\includegraphics[width=6cm]{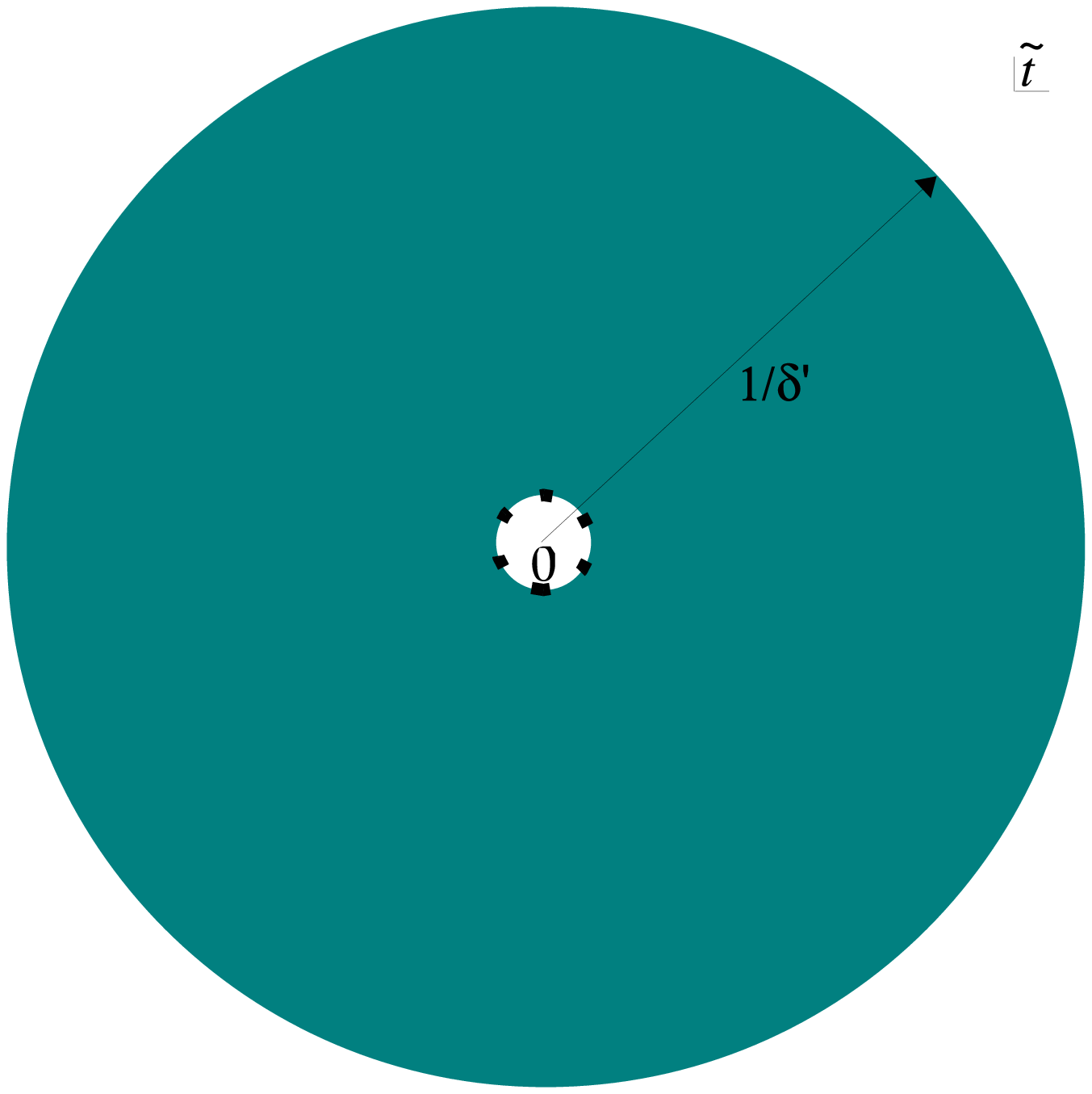}}
\caption{This figure shows how the base space, where the correlator is 
defined, gets mapped to the covering space. The ``first half'' of the
$z$-sphere is shown in grey. The holes in the $t$-sphere get patched
with flat pieces. Relative sizes, shapes, and positions are not
accurate.}\label{fig:map-regions}
\end{center}
\end{figure}

Note that there is a ring of curvature at $|z|=1/\delta$. The base
space with holes is pictured in Figure~\ref{fig:map-regions}.

We map to a covering space with coordinates $t$ and $\bar{t}$ using
the map given in Equation~\eqref{eq:themap}. The manifold still has
holes from the images of the twist operators. We close all of the
holes in the manifold by pasting in flat patches. Then, the covering
space manifold, which we call $\Sigma$, becomes compact with the
topology of a sphere\footnote{For the specific correlator we compute,
the Riemann--Hurwitz formula determines that the covering space has
genus zero. In other cases, one can get higher genus covering spaces,
as discussed in~\cite{lm1}.}. The $l+1$ $\kappa$-twist operators which
corresponded to a single hole in the base space at $\tilde{z}=0$ map
to $l+1$ different holes in the covering space, $l$ in the finite $t$
plane and one at infinity. These holes we also close with flat
patches. The covering space is pictured in
Figure~\ref{fig:map-regions}.

The metric induced on the manifold $\Sigma$ from the base metric
\bref{Eqn:BaseMetric} is conformally related to the fiducial metric
for the $t$-sphere,
\begin{equation} 
\begin{split}
\widehat{ds}^2 &= \left\{ 
\begin{array}{rl} 
dtd\bar{t} & \qquad |t| <\frac{1}{\delta'}\\
\\
d\tilde{t}d\bar{\tilde{t}} & \qquad |\tilde{t}| <\frac{1}{\delta'}\\
\end{array} \right. \\
t &= \frac{1}{{\delta'}^2\tilde{t}}.
\end{split}  \label{Eqn:FiducialMetric}
\end{equation}
We choose $\delta'$ such that the outermost image of
$|z|=\frac{1}{\delta}$ is contained in the first half of the
$t$-sphere.

Note that unlike the calculations performed in~\cite{lm1}, for our map
all $(l+1)$ images of infinity correspond to a distinct twist operator
inserted at infinity in the base space.

The Liouville field $\phi$ is defined by
\begin{equation}
ds^2 = e^\phi\,\widehat{ds}^2.
\end{equation}
We break the manifold $\Sigma$ into a ``regular region'', which is the
image of the ``first half'' of the $z$-sphere with the holes cut out;
the ``annuli'', which consist of the images of the second half of the
$z$-sphere with the hole around $\tilde{z}=0$ cut out; the ``second
half'' of the $t$-sphere with the hole cut out; and finally, the flat
patches which we pasted in to make the manifold compact. Using this
categorization, we can write $\phi$ as
\begin{equation}\label{eq:def-phi}
\phi = \begin{cases}
\log\left|\der{z}{t}\right|^2;  & \text{regular region}\\
\log\left|\der{\tilde{z}}{t}\right|^2; & \text{annuli}\\
\log\left|\der{\tilde{z}}{\tilde{t}}\right|^2; & \text{second half of the $t$-sphere}\\
\text{(constant)}; \hspace{30pt}& \text{flat patches}
\end{cases}.
\end{equation}
Note that $\phi$ is a continuous function over $\Sigma$, but its first
derivative is discontinuous across the boundaries between the above
regions.

The goal of this appendix is to compute the Liouville action,
\begin{equation}
S_L = \frac{c}{96\pi}\int_{\hat{\Sigma}}\drm^2 t\sqrt{\hat{g}}
\left[\pd_\mu\phi\pd_\nu\phi\hat{g}^{\mu\nu} + 2 \hat{R}\phi\right],
\end{equation}
in the limit of small $\veps$, $\tilde{\veps}$, $\delta$, and
$\delta'$. Any regularization dependence drops out once we normalize
the twist operators to have unit two-point function with themselves at
unit separation. It is important to note that the curvature and metric
in the above equation are on the fiducial manifold $\hat{\Sigma}$, and
not the more complicated curved manifold $\Sigma$. This means, for
instance, the only curvature contribution comes from the ring of
concentrated curvature at $|t|=1/\delta'$ where the two discs of the
fiducial metric are glued together.

There are the following nonzero contributions to the Liouville action:
\begin{enumerate}
\item a contribution $S_L^{(1)}$ from the regular region $R$ 
	(light gray in the Figure~\ref{fig:map-regions}).
\item a contribution $S_L^{(2)}$ from the annulus $A$, between 
	the outer image of $|z|=\frac{1}{\delta}$ and $|t|=1/\delta'$.
\item a contribution $S_L^{(3)}$ from the ring of curvature 
	concentrated at $|t|=1/\delta'$.
\item a contribution $S_L^{(4)}$ from the second half of the $t$-sphere,
	 but outside of the patched hole around $\tilde{t}=0$.
\item a contribution $S_L^{(5)}$ from the annuli around the finite 
	images of infinity between the image of $|\tilde{z}|=\tilde{\veps}$ 
	and the image of $|z|=|\tilde{z}|=1/\delta$.
\end{enumerate}
The flat patches (and the various boundaries between the regions) do
not contribute. In the flat regions, there is no curvature of the
fiducial metric and $\phi$ is constant; therefore, there can be no
contribution. On their boundaries, $\pd\phi$ is nonzero, but bounded
and is integrated over a set of measure zero. Therefore, the edges of
the filled holes do not contribute either. There can be no curvature
contribution beyond $S_L^{(3)}$, but one may worry about possible
kinetic term contributions from various boundaries; however, these
cannot contribute as long as $\pd\phi$ is bounded, which means that as
long as $\phi$ is continuous, there is no contribution from the
kinetic term on boundaries. Note that the fourth and fifth
contributions are zero for the calculations performed in~\cite{lm1}.

\subsection{The $S_L^{(1)}$ contributions}

The regularization-independent contributions all come from the regular
region, whose contribution is called $S_L^{(1)}$. We
have
\begin{equation}
S_L^{(1)} = \frac{c}{96\pi}\int_{\Sigma_R}\drm^2 t\sqrt{\hat{g}}\big[
\pd_\mu\phi\pd_\nu\phi\hat{g}^{\mu\nu} + 2 \hat{R}\phi\big],
\end{equation}
where the hats denote quantities computed with the fiducial metric
\begin{equation}
\widehat{ds}^2 = dtd\bar{t}\qquad |t|<\frac{1}{\delta'}.
\end{equation}
From Equation~\eqref{eq:def-phi}, one sees that in the regular region,
\begin{equation}
\phi = \log \der{z}{t} + \log \der{\bar{z}}{\bar{t}}.
\end{equation}

The fiducial metric has no curvature in the regular region $\Sigma_R$,
and therefore
\begin{equation}
S_L^{(1)} = \frac{c}{96\pi}\int_{\Sigma_R}\drm^2t\, \pd_\mu\phi\pd^\mu\phi,
\end{equation}
which we can integrate by parts into
\begin{equation}
S_L^{(1)} = -\frac{c}{96\pi}\int_{\Sigma_R}\drm^2t\,\phi\pd_\mu\pd^\mu\phi
	    +\frac{c}{96\pi}\int_{\pd\Sigma_R}\phi \pd_n\phi\, \drm s
\end{equation}
Since $\phi$ is the sum of a holomorphic part and an antiholomorphic part,
\begin{equation}
\pd_\mu\pd^\mu\phi = 4\pd\pdb\phi = 0,
\end{equation}
and therefore, we have only the boundary term to contend with. More
explicitly, one may write
\begin{equation}
S_L^{(1)} = \frac{c}{96\pi}\left[i\int_{\pd\Sigma_R}\drm t\,\phi\pd\phi 
	+ \mathrm{c.c.}\right].
\end{equation}
Note that this equation assumes a convention where ``internal''
boundaries are integrated counter-clockwise and the ``external''
boundary is integrated clockwise. Alternatively, one may think of this
prescription as integrating every contour counter-clockwise from the
local perspective of someone living inside the hole (outside the
region $\Sigma_R$) on the Riemann sphere. This means that the
``external'' hole should be viewed from $\infty$/North pole of the
Riemann sphere, which switches the contour's orientation from the
perspective of someone living in the finite $t$ plane.

The boundary of $\Sigma_R$ is composed of
\begin{enumerate}
\item the hole around the origin from the twist of order $\kappa(l+1)$.
\item the hole around $t=1$ from the twist of order $(l+1)$ at $z_2$.
\item the $l$ images of $|z|=1/\delta$ in the finite $t$ plane.
\item the outer boundary from the image of $|z|=\frac{1}{\delta}$.
\end{enumerate}
Note that each of the $(l+1)$ images of infinity corresponds to one of
the $(l+1)$ $\sigma_\kappa$ twists inserted at infinity. All contours
should be counter-clockwise, except for the contour circling
$t\to\infty$.

We begin by finding the contribution from the hole around the
origin. We make this calculation most explicit. Near the origin
\[
z \approx (-1)^{\kappa l} z_2 t^{\kappa (l+1)}\qquad 
\der{z}{t}\approx (-1)^{\kappa l}\kappa(l+1) z_2t^{\kappa(l+1) -1},
\]
and therefore (from here on we implicitly work in the limit of small
$\veps, \tilde{\veps}, \delta, \delta'$ and drop approximation signs)
\begin{equation}\begin{split}
\phi &= 2 \log\big(\kappa(l+1)\big) + 2\log|z_2| + 2\big[\kappa(l+1) -1\big]\log |t|\\
\pd\phi &= \frac{\kappa(l+1) -1}{t}.
\end{split}\end{equation}
Since the hole is of size $\veps$ in the $z$-plane, the contour in the
$t$-space should have
\begin{equation}
|t| = \left(\frac{\veps}{|z_2|}\right)^{\frac{1}{\kappa(l+1)}}.
\end{equation}
Therefore, we parameterize the contour as
\begin{equation}
t = \left(\frac{\veps}{|z_2|}\right)^{\frac{1}{\kappa(l+1)}} e^{i\theta}\qquad
\drm t = it\,\drm\theta.
\end{equation}
Thus,
\begin{calc}
S_L^{(1)}(t=0) &= \frac{c}{96\pi}\left\{-2\big[\kappa(l+1) -1\big]
	\int_0^{2\pi}\drm\theta\,\left[\log\big(\kappa(l+1)\big) + \log|z_2| 
		+ \frac{\kappa(l+1)-1}{\kappa(l+1)}\log\frac{\veps}{|z_2|}\right]\right\}\\
 &\hspace{60pt} + \mathrm{c.c}\\
 &= -\frac{c}{12}\big(\kappa(l+1) -1\big)
	\left[\frac{1}{\kappa(l+1)}\log|z_2| + \log \kappa(l+1) 
		+ \frac{\kappa(l+1)-1}{\kappa(l+1)}\log\veps\right].
\end{calc}

For $t=1$, since
\[
z\approx z_2 + z_2 \kappa (t-1)^{l+1}\qquad 
\der{z}{t}\approx z_2 \kappa(l+1) (t-1)^l,
\]
we have
\begin{equation}\begin{split}
\phi &= 2\log|z_2| + 2\log\kappa(l+1) + 2l\log|t-1|\\
\pd\phi &= \frac{l}{t-1}\\
t &= 1+ \left(\frac{\veps}{\kappa|z_2|}\right)^{\frac{1}{l+1}} e^{i\theta}\qquad
\drm t = i(t-1)\drm\theta.
\end{split}\end{equation}
Plugging in, one finds
\begin{equation}
S_L^{(1)}(t=1) = -\frac{c}{12}l\left[\frac{1}{l+1}\log|z_2| 
		+ \frac{1}{l+1}\log\left(\kappa(l+1)^{l+1}\right)
		+\frac{l}{l+1}\log\veps\right].
\end{equation}

The only difference for $t=\infty$ is that the contour is clockwise
which gives an extra minus sign. From above, we know that
\[
z \approx \frac{z_2}{(l+1)^\kappa}t^\kappa\qquad
\der{z}{t} \approx \frac{z_2\kappa}{(l+1)^\kappa} t^{\kappa -1},
\]
and therefore
\begin{equation}\begin{split}
\phi &= 2\log|z_2| + 2\log\frac{\kappa}{(l+1)^\kappa} + 2(\kappa -1)\log|t|\\
\pd\phi &= \frac{\kappa -1}{t}\\
t &= \frac{l+1}{(\delta|z_2|)^{\frac{1}{\kappa}}}e^{-i\theta}\qquad
\drm t = -it\drm \theta.
\end{split}\end{equation}
Thus,
\begin{equation}
S_L^{(1)}(t\to\infty) = \frac{c}{12}(\kappa -1)
	\left[\frac{1}{\kappa}\log|z_2|
		+ \log \frac{\kappa}{l+1} - \frac{\kappa -1}{\kappa}\log\delta\right].
\end{equation}

Finally, the only other contribution to $S_L^{(1)}$ comes from the
images of $z\to\infty$ that are in the finite $t$-plane. Near one of
the $t_\infty^{(j)},$\footnote{we dropped some irrelevant minus signs
in the map.}
\[
z \approx z_2\left[\frac{\alpha_j}{(l+1)(1-\alpha_j)^2}\right]^\kappa
	\frac{1}{(t-t_\infty^{(j)})^\kappa}\qquad
\der{z}{t}\approx -\kappa z_2 \left[\frac{\alpha_j}{(l+1)(1-\alpha_j)^2}\right]^\kappa
	\frac{1}{(t-t_\infty^{(j)})^{\kappa+1}},
\]
where recall that the
\[
t_\infty^{(j)} = \frac{1}{1-\alpha_j}.
\]
For convenience, we define
\begin{equation}
\beta_j = \left[\frac{\alpha_j}{(l+1)(1-\alpha_j)^2}\right]^\kappa
	= \frac{1}{\big[(l+1)\alpha_j^l(1-\alpha_j)^2\big]^\kappa}.
\end{equation}
The above yields
\begin{equation}\begin{split}
\phi &=  2\log|z_2| + 2\log \kappa|\beta_j| -2(\kappa + 1)\log |t-t_\infty^{(j)}|\\
\pd\phi &= -\frac{\kappa +1}{t-t_\infty^{(j)}}\\
t &= t_\infty^{(j)} +\big(|z_2| |\beta_j|\delta\big)^\frac{1}{\kappa}e^{i\theta}\qquad
\drm t = i(t-t_\infty^{(j)})\drm \theta.
\end{split}\end{equation}
Being careful of minus signs, one finds
\begin{equation}
S_L^{(1)}(t=t^{(j)}) = -\frac{c}{12}(\kappa+1)
	\left[\frac{1}{\kappa}\log|z_2| - \log\kappa + \frac{1}{\kappa}\log|\beta_j|
		+\frac{\kappa+1}{\kappa}\log\delta\right].
\end{equation}
Adding all $l$ contributions from finite images of infinity,
\begin{equation}
\sum_{j=1}^l S_L^{(1)}(t^{(j)}) = -\frac{c}{12}(\kappa+1)l
	\left[\frac{1}{\kappa}\log|z_2| - \log\kappa 
	      +\frac{\kappa+1}{\kappa}\log\delta\right]
	-\frac{c}{12}\frac{\kappa+1}{\kappa}\sum_{j=1}^l\log|\beta_j|.
\end{equation}
To simplify the final term note that
\begin{equation}
\sum_{j=1}^l\log|\beta_j| = -\kappa\sum_{j=1}^l
	\log \left[(l+1)|\alpha_j^l(1-\alpha_j)^2|\right]
	= -\kappa l \log (l+1) -\kappa
	\log \left|\prod_{j=1}^l \alpha_j^{l}(1-\alpha_j)^2\right|.
\end{equation}
Thus, we become interested in computing
\begin{calc}
\prod_{j=1}^l\alpha_j^l (1-\alpha_j)^2 &= \frac{\left(\prod_{j=1}^l(1-\alpha_j)\right)^2}
	{\prod_{j=1}^l\alpha_j}\\
	&= (l+1)^2,
\end{calc}
where we have used (from~\cite{lm1} Equation~(3.16))\footnote{There is
  an error in~\cite{lm1}, which does not affect the final answer.}
\begin{equation}
\prod_{j=1}^l \alpha_j =(-1)^l\qquad
\prod_{j=1}^l (q-\alpha_j) = \frac{q^{l+1}-1}{q-1}\xrightarrow[q\to 1]{}l+1.
\end{equation}
Therefore, the total contribution from the finite images of infinity
is
\begin{calc}
\sum_{j=1}^l S_L^{(1)}(t^{(j)}_\infty) &= -\frac{c}{12}(\kappa+1)l
	\left[\frac{1}{\kappa}\log|z_2| - \log\kappa 
	      +\frac{\kappa+1}{\kappa}\log\delta\right]
	+ \frac{c}{12}(\kappa +1)(l+2)\log (l+1)\\
	&= -\frac{c}{12}(\kappa+1)l
	\left[\frac{1}{\kappa}\log|z_2| - \log\kappa -\frac{l+2}{l}\log(l+1)
	      +\frac{\kappa+1}{\kappa}\log\delta\right].
\end{calc}

Summing up all of the contributions, we find that
\begin{equation}\begin{split}
S_L^{(1)} = -\frac{c}{12}\biggr\{&\frac{(\kappa+1)l(l+2)}{\kappa(l+1)}\log|z_2|\\
	& - \frac{l^2}{l+1}\log\kappa\\
	&-4\log(l+1)\\
	&+\frac{1}{l+1}\left(l^2(\kappa+1)+2l(\kappa -1) 
		+ \frac{(\kappa-1)^2}{\kappa}\right)\log\veps\\
	&+\left(\frac{l+1}{\kappa}(\kappa^2+1)+2(l-1)\right)\log\delta\biggr\}.
\end{split}\end{equation}
Let's examine the $z_2$ dependence. The expected power of $|z_2|$ for
the correlator is
\begin{calc}
-2\left(\Delta_{\kappa(l+1)}+\Delta_{l+1} - (l+1)\Delta_{\kappa}\right)
	&= -\frac{c}{12}\bigg[\kappa(l+1) - \frac{1}{\kappa(l+1)}
		+(l+1) -\frac{1}{l+1}\\ 
	&\hspace{143pt}-(l+1)\left(\kappa-\frac{1}{\kappa}\right)\bigg]\\
	&= -\frac{c}{12}\frac{(\kappa+1)l(l+2)}{\kappa(l+1)}.
\end{calc}
This exactly matches the contribution from $S_L^{(1)}$. Therefore, any
other contributions' $|z_2|$-dependence should cancel out.

\subsection{The contribution $S_L^{(2)}$}

We now compute the contribution to the Liouville action from the outer
annulus between the image of $|z|=1/\delta$ and $|t|=1/\delta'$.  From
the behavior of the map for large $t$,
\[
z \approx \frac{z_2}{(l+1)^\kappa}t^\kappa
\]
we see that the region for this contribution is defined by
\begin{equation}
\frac{l+1}{\big(|z_2|\delta\big)^\frac{1}{\kappa}}
< |t|<
\frac{1}{\delta'}.
\end{equation}
The induced metric for this region comes from the second half of the
$z$-plane:
\begin{equation}
ds^2 = d\tilde{z}d\bar{\tilde{z}} = 
  \der{\tilde{z}}{t}\der{\bar{\tilde{z}}}{\bar{t}}dtd\bar{t}
   = e^\phi dtd\bar{t}.
\end{equation}
Thus, we should look at 
\begin{equation}\begin{split}
\tilde{z} &= \frac{1}{\delta^2}\frac{(l+1)^\kappa}{z_2 t^\kappa}\qquad
\der{\tilde{z}}{t} = -\frac{\kappa}{\delta^2}
                      \frac{(l+1)^\kappa}{z_2 t^{\kappa+1}}\\
\phi &= 2\big[\log\kappa(l+1)^\kappa-2\log\delta 
              -\log|z_2|-(\kappa+1)\log|t|\big]\\
\pd\phi &= -\frac{\kappa+1}{t}.
\end{split}\end{equation}
One can integrate by parts, or integrate directly, as we do below:
\begin{calc}
S_L^{(2)} &= \frac{4c}{96\pi}\int\drm^2 t\, \pd\phi\pdb\phi\\
          &= \frac{c}{24\pi} \int\big(2\pi|t|\drm|t|\big)\frac{(\kappa+1)^2}{|t|^2}\\
          &= \frac{c}{12}(\kappa+1)^2\int\frac{\drm|t|}{|t|}\\
          &= \frac{c}{12}\frac{(\kappa+1)^2}{\kappa}
                \log\frac{|z_2|\delta}{\big((l+1)\delta'\big)^\kappa}.
\end{calc}

\subsection{The contribution $S_L^{(3)}$}

Here we are just concerned with the curvature term. Since all of the
curvature is concentrated along the ring $|t|=1/\delta'$, we find
\begin{equation}
S_L^{(3)} = \frac{c}{48\pi}\int\drm^2t\, \hat{R}\phi 
          = \frac{c}{6}\,\phi\big|_{|t|=\frac{1}{\delta'}}.
\end{equation}
From above, we see that
\begin{calc}
\phi &= 2\big[\log\kappa(l+1)^\kappa-2\log\delta 
              -\log|z_2|-(\kappa+1)\log|t|\big]\\
     &= -2\left[\log|z_2|-\log\kappa(l+1)^\kappa 
                + \log\frac{\delta^2}{\delta'^{\kappa+1}}
          \right].
\end{calc}
Plugging in, one finds
\begin{equation}
S_L^{(3)} = -\frac{c}{3}\left[\log|z_2|-\log\kappa(l+1)^\kappa 
                + \log\frac{\delta^2}{\delta'^{\kappa+1}}
          \right].
\end{equation}

\subsection{The contribution $S_L^{(4)}$}

We now calculate our first new contribution. We should find that it
vanishes for $\kappa=1$. The map in this region is between $\tilde{z}$
and $\tilde{t}$:
\begin{equation}\begin{split}
\tilde{z} &= \left(\frac{\delta'^\kappa}{\delta}\right)^2
            \frac{(l+1)^\kappa}{z_2}\tilde{t}^\kappa\qquad
\der{\tilde{z}}{\tilde{t}}
   =\left(\frac{\delta'^\kappa}{\delta}\right)^2
            \frac{\kappa(l+1)^\kappa}{z_2}\tilde{t}^{\kappa-1}\\
\phi &= 2\left[\log\kappa(l+1)^\kappa - \log|z_2| 
               + 2\log\frac{{\delta'}^\kappa}{\delta} 
               + (\kappa - 1)\log|\tilde{t}|\right]\\
\pd\phi &= \frac{\kappa-1}{\tilde{t}}.
\end{split}\end{equation}
As for $S_L^{(2)}$, it may be more convenient to integrate directly
over the region,
\begin{equation}
\frac{1}{l+1}\left(\tilde{\veps}\frac{\delta^2}{{\delta'}^{2\kappa}}
                   |z_2|\right)^\frac{1}{\kappa}
< |\tilde{t}| < \frac{1}{\delta'}.
\end{equation}
Proceeding, one finds
\begin{calc}
S_L^{(4)} &= \frac{c}{12}(\kappa-1)^2\int\frac{\drm|t|}{|t|}\\
          &= \frac{c}{12}\frac{(\kappa-1)^2}{\kappa}
   \log \frac{(l+1)^\kappa{\delta'}^{\kappa}}
             {|z_2|\tilde{\veps}\delta^2}.
\end{calc}

\subsection{The contribution $S_L^{(5)}$}

Finally, all that remains is the contribution from the annuli
surrounding the finite images of infinity, where the induced metric
comes from the second half of the $z$-sphere. Hence, the map is
between $\tilde{z}$ and $t$:
\begin{equation}
z = z_2 \beta_j\frac{1}{(t-t_\infty^{(j)})^\kappa}\Longrightarrow
\tilde{z} = \frac{1}{\delta^2}\frac{(t-t_\infty)^\kappa}{z_2\beta},
\end{equation}
where we have given the map near the $j$th finite image of infinity.
Note that we dropped the index $j$, since it can be shown that all of the annuli
give the same contribution.

The region, of interest is 
\begin{equation}
\big(\delta^2\tilde{\veps}|z_2||\beta|\big)^\frac{1}{\kappa}
< |t-t_\infty|
< \big(\delta|z_2| |\beta|\big)^\frac{1}{\kappa}.
\end{equation}

Once again, we find it most convenient to integrate directly, in which
case, we need only know
\begin{equation}
\pd\phi = \frac{\kappa -1}{t-t_\infty}.
\end{equation}
Integrating, one produces the expression for the contribution from
\emph{one} of the annuli:
\begin{equation}
S_L^{(5)}(t_\infty^{(j)}) = -\frac{c}{12}\frac{(\kappa -1)^2}{\kappa}
                             \log(\tilde{\veps}\delta).
\end{equation}
As claimed, the contribution did not depend on which finite image of
infinity under consideration; therefore, one computes the total
contribution of this kind by multiplying the above expression by $l$
to find
\begin{equation}
S_L^{(5)} = -\frac{c}{12}\frac{l(\kappa -1)^2}{\kappa}
                             \log(\tilde{\veps}\delta).
\end{equation}

\subsection{Adding up all of the contributions}

For convenience, we collect all of the contributions:
\begin{subequations}
\begin{align}
&\begin{aligned}
S_L^{(1)} &= -\frac{c}{12}\biggr\{\frac{(\kappa+1)l(l+2)}{\kappa(l+1)}\log|z_2|\\
	&\hspace{50pt} - \frac{l^2}{l+1}\log\kappa\\
	&\hspace{50pt}-4\log(l+1)\\
	&\hspace{50pt}+\frac{1}{l+1}\left(l^2(\kappa+1)+2l(\kappa -1) 
		+ \frac{(\kappa-1)^2}{\kappa}\right)\log\veps\\
	&\hspace{50pt}+\left(\frac{l+1}{\kappa}(\kappa^2+1)+2(l-1)\right)\log\delta\biggr\}\\
\end{aligned}\\
&S_L^{(2)} = \frac{c}{12}\frac{(\kappa+1)^2}{\kappa}
                \log\frac{|z_2|\delta}{\big((l+1)\delta'\big)^\kappa}\\
&S_L^{(3)} = -\frac{c}{3}\left[\log|z_2|-\log\kappa(l+1)^\kappa 
                + \log\frac{\delta^2}{\delta'^{\kappa+1}}
          \right]\\
&S_L^{(4)} = \frac{c}{12}\frac{(\kappa-1)^2}{\kappa}
   \log \frac{(l+1)^\kappa{\delta'}^{\kappa}}
             {|z_2|\tilde{\veps}\delta^2}\\
&S_L^{(5)} = -\frac{c}{12}\frac{l(\kappa -1)^2}{\kappa}
                             \log(\tilde{\veps}\delta).
\end{align}
\end{subequations}

\begin{subequations}
Note that the $|z_2|$ dependence of $S_L^{(2)}$ through $S_L^{(5)}$
cancels out, leaving us, correctly, with just the dependence of
$S_L^{(1)}$. Another check is to compute the $\delta'$ dependence,
\begin{equation}
\delta':\qquad \frac{c}{3}\log\delta',
\end{equation}
which is exactly the correct dependence to get cancelled by
$Z_{\delta'}$ in Equation~\eqref{eq:def-corr}. We discuss this in more
detail in Section~\ref{sec:norm}. Computing the $\tilde{\veps}$
dependence, one finds
\begin{equation}
\tilde{\veps}: \qquad -\frac{c}{12}\frac{(l+1)(\kappa-1)^2}{\kappa}\log\tilde{\veps},
\end{equation}
which should get canceled out by the normalization of the two-point
function. The $\delta$ dependence is a bit more tedious to work out,
but comes to
\begin{equation}
\delta:\qquad -\frac{c}{12}\frac{2(l+1)(\kappa^2+1)}{\kappa}\log\delta.
\end{equation}
Finally, there is the regularization-independent part, about which we
cannot say anything until we normalize the twists in
Section~\ref{sec:norm}:
\begin{equation}\begin{split}
\kappa: \qquad &\frac{c}{12}\frac{(l+2)^2}{l+1}\log\kappa\\
(l+1):\qquad   &\frac{c}{3}\log(l+1).
\end{split}\end{equation}
\end{subequations}
Putting all of the pieces together, we have the expression for the
total Liouville action:
\begin{equation}\begin{split}\label{eq:SL-tot}
S_L^{\text{tot.}} = -\frac{c}{12}\biggr\{&\frac{(\kappa+1)l(l+2)}{\kappa(l+1)}\log|z_2|\\
                              &-\frac{(l+2)^2}{l+1}\log\kappa
                               -4\log(l+1)\\
                              &+\frac{1}{l+1}\left(l^2(\kappa+1)+2l(\kappa -1) 
		+ \frac{(\kappa-1)^2}{\kappa}\right)\log\veps\\
                              &+\frac{(l+1)(\kappa-1)^2}{\kappa}\log\tilde{\veps}\\
                              &+\frac{2(l+1)(\kappa^2+1)}{\kappa}\log\delta\\
                              &-4\log\delta'\biggr\}.
\end{split}\end{equation}

\section{The two-point function for twists at infinity}\label{ap:two-point}

In order to normalize the correlator, we need to compute the two-point
function of $\sigma_\kappa$ at the origin with $\sigma_\kappa$ at
infinity using the same regularizations as in
Appendix~\ref{sec:liouville-action}. The map is simply
\begin{equation}
z = b t^\kappa.
\end{equation}
To find the total Liouville action for this map, we recognize that
this case is given by setting $l=0$ in the calculation of
Appendix~\ref{sec:liouville-action}:
\begin{equation}
S_L^{\mathrm{tot.}} = -\frac{c}{12}\left\{
-4\log\kappa
+\frac{(\kappa-1)^2}{\kappa}\log\veps
+\frac{(\kappa-1)^2}{\kappa}\log\tilde{\veps}
+2\frac{\kappa^2+1}{\kappa}\log\delta
-4\log\delta'\right\}.
\end{equation}
Note that the $b$ dependence cancelled out, as it must.

Since the two-point correlator is given by 
\[
\vev{\sigma_\kappa(\infty)\sigma_\kappa(0)} = e^{S_L}\frac{Z_{\delta'}}{(Z_\delta)^\kappa},
\]
and
\begin{equation}
Z_\delta = Q\delta^{-\frac{c}{3}}\qquad
Z_{\delta'} = Q{\delta'}^{-\frac{c}{3}},
\end{equation}
the two-point function is given by
\begin{equation}\label{eq:two-pt-infty}
\vev{\sigma_\kappa(\infty)^{\tilde{\veps},\delta}\sigma_\kappa(0)^\veps}_\delta
 = \frac{\kappa^{\frac{c}{3}}\delta^{\frac{c}{6}\frac{\kappa^2-1}{\kappa}}}{Q^{\kappa-1}
	(\veps\tilde{\veps})^{\frac{c}{12}\frac{(\kappa-1)^2}{\kappa}}}.
\end{equation}

Note that if we attempted to normalize the two twists using
Equation~\eqref{eq:normalized-twists} (with respect to their local
appearance), then one finds the correlator of the normalized twists is
\begin{equation}
\vev{\sigma_\kappa(\infty)\sigma_\kappa(0)}_\delta = \delta^{\frac{c}{6}\frac{\kappa^2-1}{\kappa}}
						   = \delta^{4\Delta_\kappa},
\end{equation}
exactly what one would expect for the two-point function of two
normalized twists with one at the origin and one at $|z|=1/\delta$.

\bibliography{fuzzball-decays}

\end{document}